\documentclass[nologo,10pt,a4paper]{ETHpaper}

\usepackage{verbatim}  \usepackage{graphicx}                  \usepackage[compress]{natbib}     
\usepackage{amsfonts}
\usepackage{array}
\usepackage{amsthm}
\usepackage{amsmath}
\usepackage{caption}
\usepackage{soul}
\graphicspath{{fig/}}

\usepackage{color}

\newcommand{\mean}[1]{\left\langle #1 \right\rangle}

\renewcommand{\epsilon}{\varepsilon} 
 
\renewcommand*{\=}{{\kern0.1em=\kern0.1em}}
\renewcommand*{\-}{{\kern0.1em-\kern0.1em}} 
\newcommand*{\+}{{\kern0.1em+\kern0.1em}}

\begin{document}

\title{Data-driven modeling of collaboration networks:\\  A cross-domain analysis}

\titlealternative{Data-driven modeling of collaboration networks:  A cross-domain analysis}

\author{Mario V. Tomasello,$^{1,2}$ Giacomo Vaccario$^{1,*}$, Frank Schweitzer$^1$}

\address{
$^*$Corresponding author; E-mail: gvaccario@ethz.ch\\
$^1$Chair of Systems Design, ETH Zurich, 
Weinbergstrasse 56/58, CH-8092 Zurich, Switzerland\\
$^{2}$current: EDGE Strategy, Vorstadt 2, CH-6300 Zug, Switzerland
}

\authoralternative{M. V. Tomasello, G. Vaccario, F. Schweitzer}

\reference{(submitted for publication)}
\maketitle

\begin{abstract}
We analyze large-scale data sets about collaborations from two different domains: economics, specifically 22.000 R\&D alliances between 14.500 firms,  and science, specifically 300.000 co-authorship relations between 95.000 scientists. 
Considering the different domains of the data sets, we address two questions: (a) to what extent do the collaboration networks reconstructed from the data share common structural features, and (b) can their structure be reproduced by the same agent-based model. 
In our data-driven modeling approach we use aggregated network data to calibrate the probabilities at which agents establish collaborations with either newcomers or established agents.   
The model is then validated by its ability to reproduce network features not used for calibration, including distributions of degrees, path lengths, local clustering coefficients and sizes of disconnected components. 
Emphasis is put on comparing domains, but also sub-domains (economic sectors, scientific specializations). 
Interpreting the link probabilities as strategies for link formation, we find that in R\&D collaborations newcomers prefer links with established agents, while in co-authorship relations newcomers prefer links with other newcomers. 
Our results shed new light on the long-standing question about the role of endogenous and exogenous factors (i.e., different information available to the initiator of a collaboration) in network formation.

\end{abstract}

\section{Introduction}
\label{sec:intro}

The availability of large-scale and time resolved data sets about economic, scientific or social activities opens new venues to address the long standing question {of} how we collaborate. 
This question becomes more important as globalization leads to a vast increase of collaborations in many areas {of human activity}, including science and economics 
\citep{narin1991globalization, luukkonen1992understanding, Georghiou1998GlobalCoop, hagedoorn2002inter}.
One could argue that collaboration patterns change with respect to the actors and the domain of activity, but there may be also evidence for common features across different domains. 
In the latter case, we could hypothesize that a unified modeling approach should be able to reproduce, and to explain, the structural and the dynamic features of collaborations in different domains.
To demonstrate this is the aim of our paper. 

{The present study is focused on two domains with a large impact on human development, (i) economy and (ii) science. 
Specifically, we refer to (i) firms collaborating in Research and Development (R\&D) alliances and (ii) scientists collaborating in co-authored publications.} 
For both cases large, comprehensive and structured data sets about individual collaboration activities have become available.
{The data sets analyzed in this study are} (i) the Thomson Reuters SDC Platinum database, listing around 15 000 inter-firm R\&D alliances and (ii) a data set of over 300 000 {co-authored papers} in physics, which was obtained from the APS scholars database with additional disambiguation of authors names. 
For the details we refer to Section \ref{sec:data_and_methodology}. 

The time-aggregated data about these collaboration events can be conveniently represented by means of a complex network, where the nodes are the actors{,} or \emph{agents} as we denote them in the following, and the links are the recorded collaborations. 
The structural features of such collaboration networks have been already {investigated} in different domains. 
Previous works have, for instance, discussed the presence of \emph{clusters}, or communities, both in R\&D networks of firms \citep{rosenkopf2008investigating,tomasello2016riseandfall} and in co-authorship networks of scientists \citep{Newman2001Clustering}. 
{The existence of such communities also impacts performance criteria} \citep{guimera2005team,Sarigol2014} and affect{knowledge transfer} \citep{tomasello2016knowledge_exchange,sorenson06:_compl} and {the ability to innovate} \citep{KoBaNaSchJEBO,sammarra2008heterogeneity,valverde07:_topol}. 
Other topological analyses focus on importance measures to characterize nodes \citep{Scholtes2016,estrada05:_subgr,borgatti05:_centr}.

However, even the most refined topological characterization of collaboration networks can only constitute a first step toward their comprehensive and systematic understanding. 
This has to include the mechanisms that shape the structure and dynamics of such networks at the level of nodes, or agents.  
In particular, we need to identify the \emph{rules}, or \emph{strategies}, that agents follow in choosing their collaboration partners -- such that at the end the observed collaboration networks emerge.

To combine the empirical analysis with a formal approach of the network formation we have proposed \emph{data-driven modeling} as a suitable methodology. 
It is, for the application at hand, comprised of the following four steps: 
(a) proposition of an \emph{agent-based model} (ABM) that shall explain the formation of collaboration networks,
(b) \emph{reconstruction} of the collaboration networks using the empirical data from two different domains,
(c) \emph{calibration} of the free parameters of the ABM for each domain by means of the empirical networks,
(d) \emph{validation} of the ABM for each domain by reproducing network features not used for the calibration. 

This leaves us with the question about agent-based models that are suitable for being used in a data-driven approach. 
Some ABM rooted in economics propose a utility function for an agent which weight costs and benefits of collaborations \citep{KoBaNaSchJEBO, konig2012efficiency}.
Agents create or maintain links only if this mutually increases their utility, and delete existing links otherwise. 
Such ABM allow to prove general features of, e.g., R\&D networks such as sparseness or stability, dependent on certain cost functions. 
But because of theoretical assumptions about the utility function and the partner selection they cannot easily be calibrated against network data. 
Therefore, we have developed an ABM in the context of R\&D collaborations \citep{tomasello2014therole} 
which assumes simple rules of link formation that are followed by agents with certain probabilities (see Section \ref{sec:agent-based-model} for details). 
Such probabilities can be calibrated against available network data. 

In this paper, we build on the existing ABM  \citep{tomasello2014therole} which was already applied to R\&D alliances \citep{tomasello2015quantifying,garas2017selection}, but has not been extended to, or validated in, other domains yet. 
Hence, the goal of this work is twofold. On the one hand, we want to understand whether the same agent-based model can reproduce the topology of both R\&D and co-authorship networks. 
On the other hand, we want to identify similarities and differences - at the \textit{microscopic} level - with respect to the agents' choice of collaboration partners.
To the best of our knowledge no study has tried yet to unify findings in these two domains and find systematic, reproducible and universal patterns in collaboration networks.
This investigation can also {provide some} evidence to our initial conjecture whether there may be a unified modeling approach for collaboration networks in different domains.

\begin{figure}[htbp]
\begin{center}
(a)\includegraphics[width=0.45\textwidth]{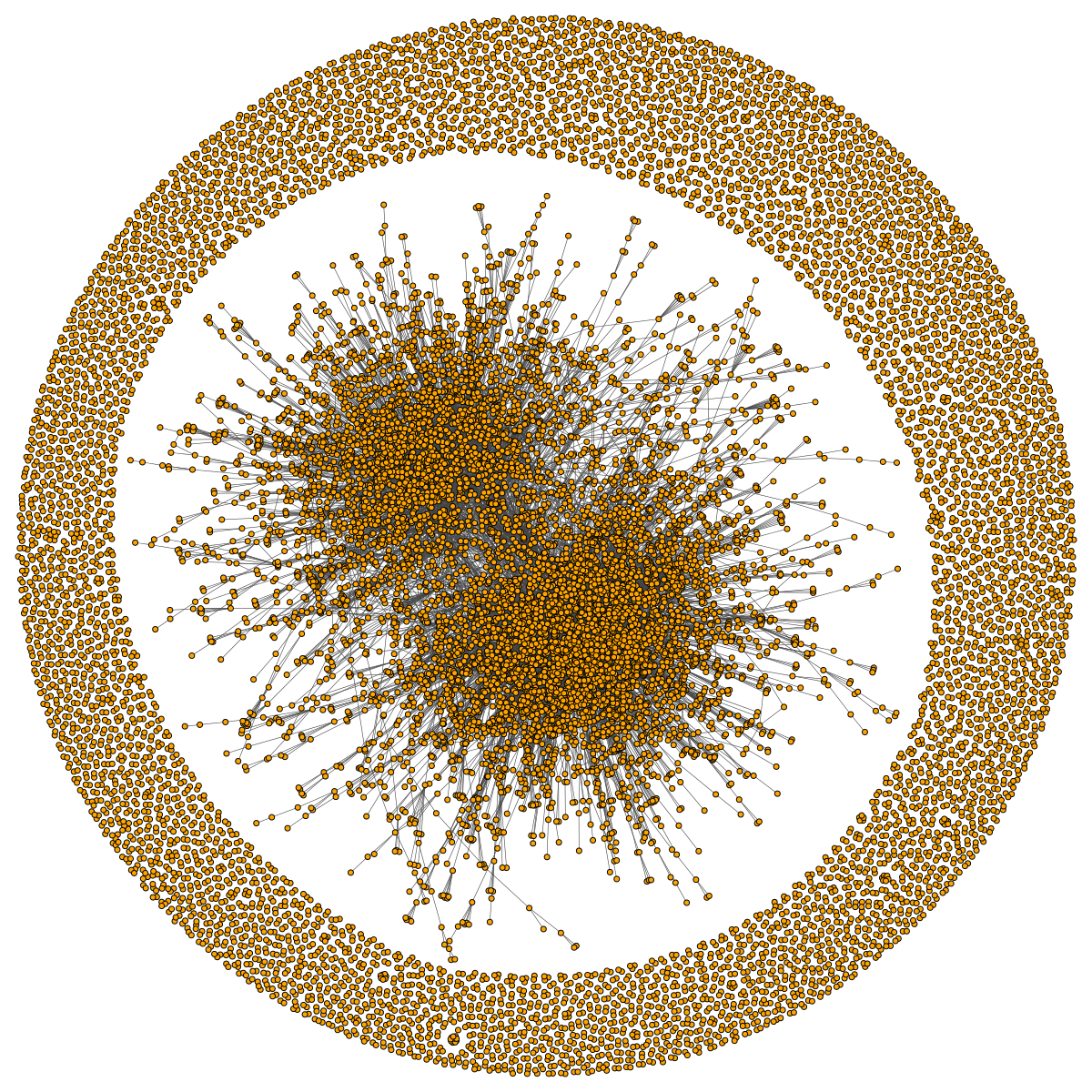}
(b)\includegraphics[width=0.45\textwidth]{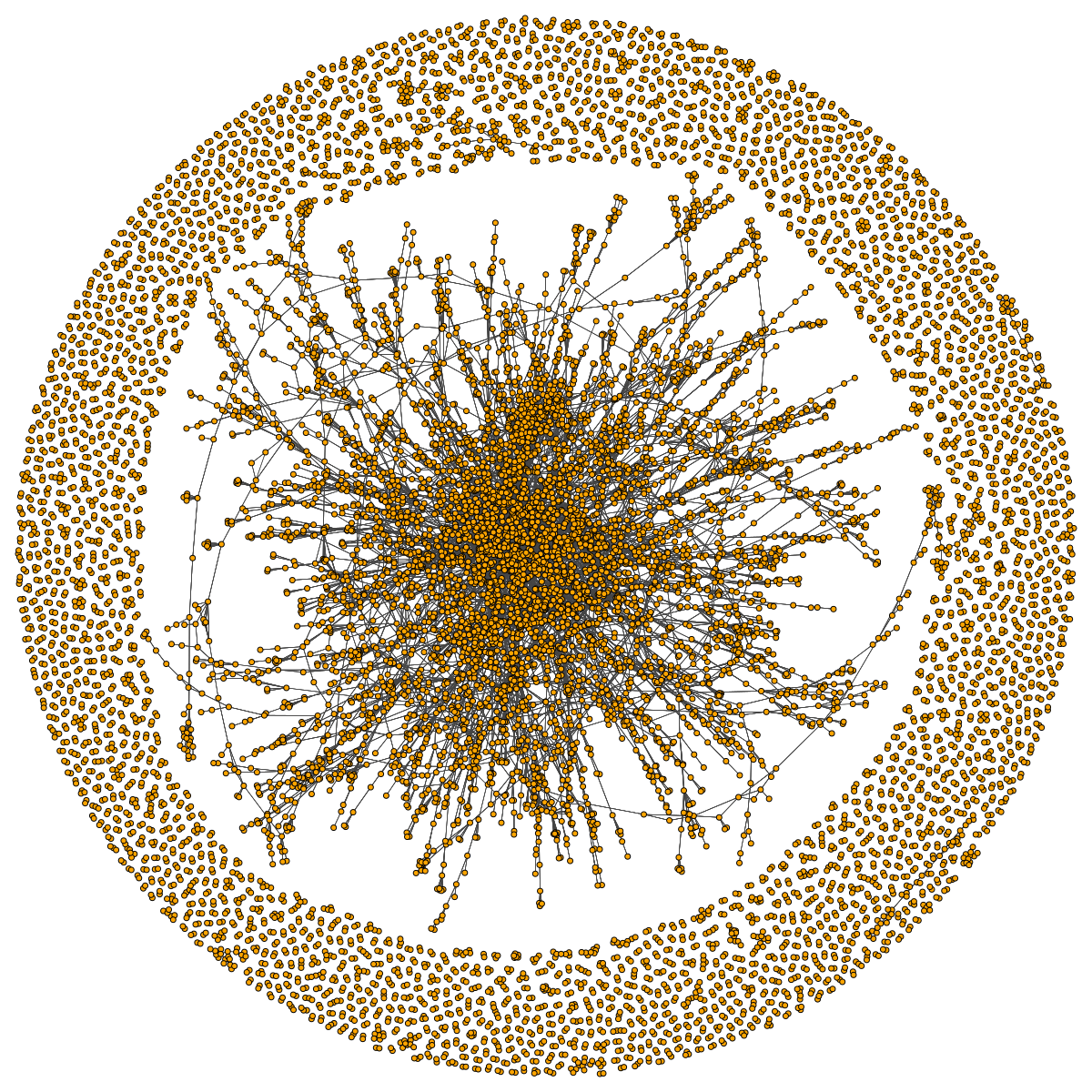} 
\end{center}
\vspace{-4pt}
\caption[]{Visualization of collaboration networks: (left) R\&D alliances of firms, (right) co-authorship relations of scientists. 
For the data sets see Section \ref{sec:data_and_methodology}. We show the complete R\&D network with about 14 000 nodes and 21 000 links, but only a sampled co-authorship network with about 11 000 nodes and 32 000 links (i.e. 10\% of randomly chosen co-authors). 
 For both networks we use the layout algorithm of \citet{fruchterman_reingold_1991}.
 }

\label{fig:network_plots}
\end{figure}

\section{Agent-based model of collaborations}
\label{sec:agent-based-model}

How do economic actors or scientists choose their collaboration partners? 
{At first, one would argue that scientists as decision makers} are quite different from firms.
In addition, {inside their respective domain,} how they choose partners may very much depend on the specific economic sector or  scientific discipline. 
Thus, there is no ad-hoc evidence that such a problem can be addressed using the same modeling framework. 

On the other hand, in order to reproduce a macroscopic structure such as a collaboration network, we may not need to include all the microscopic details that distinguish economic from social agents. 
Instead, an agent-based model should abstract from these details, to capture only the essential features of the decision making process. 
In this sense, we aim at an agent-based model that includes a minimalistic set of microscopic rules.
We argue that this agent-based model is correct if it is able to reproduce a specific set of macroscopic properties of the different collaboration networks,  namely degree distribution, path length distribution, distribution of community sizes, that {are} \emph{not} used for the calibration of the model. 
{At the same time}, the agent-based model has to provide degrees of freedom to allow a proper calibration {to reflect} the differences of the domains in their respective empirical data. 

{In order to achieve this goal, this study utilizes a} previously proposed agent-based model \citep{tomasello2014therole} that has the above mentioned features. 
{The model is flexible in that it builds on five probabilities to capture the choice of agents for collaborating with either established nodes or newcomers, which need to be calibrated.} 
Obviously, different sets of probabilities may match the same macroscopic features. 
In order to distinguish between them, we adopt a Maximum-Likelihood approach that uses the mean degree, the mean path length, and the global clustering coefficient of the resulting collaboration network as quantities to be exactly matched. 

In the model, agents represent nodes in a collaboration network and links between nodes represent collaboration events. 
Each agent is characterized by two individual attributes, activity $a_{i}$ and label $l_{i}$. 
\textit{Activity} reflects the propensity to participate in a collaboration, while \emph{label} indicates that the agent belongs to a group with a certain influence, discussed in detail below. 
The agent's dynamics can be divided in two steps: \emph{first}, the agent decides with whom to link, which impacts the network topology and the size of the network if a newcomer is chosen. 
\emph{{Second}}, she adjusts her label, i.e. she keeps her previous label if she already {has} one, or she adopts the label of the counterparty if she is a newcomer, or she receives a new label, as discussed below.

\paragraph{Activation.}

The model is initialized by assigning individual activities $a_{i}$ to agents which are sampled without replacement from the empirical 
distribution of activities (see Section \ref{sec:data_and_methodology}).
Hence, these activities are different for each agent and kept constant in time for the simulation.
Next, at each time step, we select an agent to initiate a collaboration with probability $p_{i}$ proportional to its activity,
$p_{i}= \eta a_{i}$, where $\eta$ is a rescaling parameter that we fix by imposing that $\sum_i p_{i}$ is equal to the number of collaboration event empirically observed per day.
\paragraph{{Non-labeled}
 versus labeled agents.}
Activated agents can belong to two different groups: (a) newcomers, if they never engaged in a collaboration before, or (b) established agents, if they were already part of a previous collaboration. 
We distinguish between these groups by means of the agent label $l_{i}$. 
Newcomers are non-labeled, $l_{i}=0$, whereas established agents get a label depending on their first collaboration, $l_{i}>0$. 

\paragraph{Collaboration size.}
When an agent is activated, she initiates a collaboration. 
The number of partners for her collaboration, $m_{i}$, is obtained by sampling at random from the empirical size distribution of collaborating groups (see Section \ref{sec:data_and_methodology}). 
The selection of partners is independent of the activity or other characteristics of the agent.

\paragraph{Collaboration partners.}

Given the size of the collaboration, the initiator chooses partners either from the group of newcomers or from the group of established agents. 
This choice also depends on the label of the initiator herself and can be expressed by five probabilities.
A labeled initiator links to another agent with the \emph{same} label with probability $p^{L}_{s}$, to an agent with a \emph{different} label with probability $p^{L}_{d}$, or to an agent \emph{without} any label with probability $p^{L}_{n}$.
If the initiator is a newcomer, i.e. non-labeled, she links to an \emph{labeled} agent with probability $p^{N\!L}_{l}$ and to another \emph{newcomer} with probability $p^{N\!L}_{n}$. 
Because the probabilities have to sum up to one, we have two constrains $p^{L}_{s}+p^{L}_{d}+p^{L}_{n}=1$ and $p^{N\!L}_{n}+p^{N\!L}_{l}=1$. 

\paragraph{Link formation.}

The probabilities to choose collaboration partners only consider the two groups, newcomers and established agents. 
To specify which of the specific agents from these groups are chosen, we adopt the preferential attachment rule. 
Precisely, the initiator $i$ selects, among all agents from the specific group, agent $j$ as collaborator with a probability proportional to the degree $k_j$ of $j$. If the initiator chooses a non-labeled agent ($k_j=0$) as collaborator, she will select uniformly at random from 
all non-labeled agents.
After selecting the $m_{i}$ partners, we link all of them to the initiator, this way creating a clique of size $m+1$. 
\paragraph{Label dynamics.}

In our model, agents are initialized as non-labeled agents, i.e. they are considered as newcomers. 
An agent receives a label only when entering the network (which may consist of disconnected communities).  
This can happen in two different ways: either the agent initiates a collaboration, or the agent is chosen as partner by an activated agent. 
In the first case, the agent gets a new label assigned that was not used before. 
In the second case, the agent adopts the label from the initiator of the collaboration.
The label is a unique attribute of an agent, i.e. once an agent has obtained a label, this cannot be changed. 

\bigskip 

\begin{figure}[htpb]
 \vspace{6pt}
 \begin{center}
 \footnotesize{(a)}
  \includegraphics[width=0.3\textwidth]{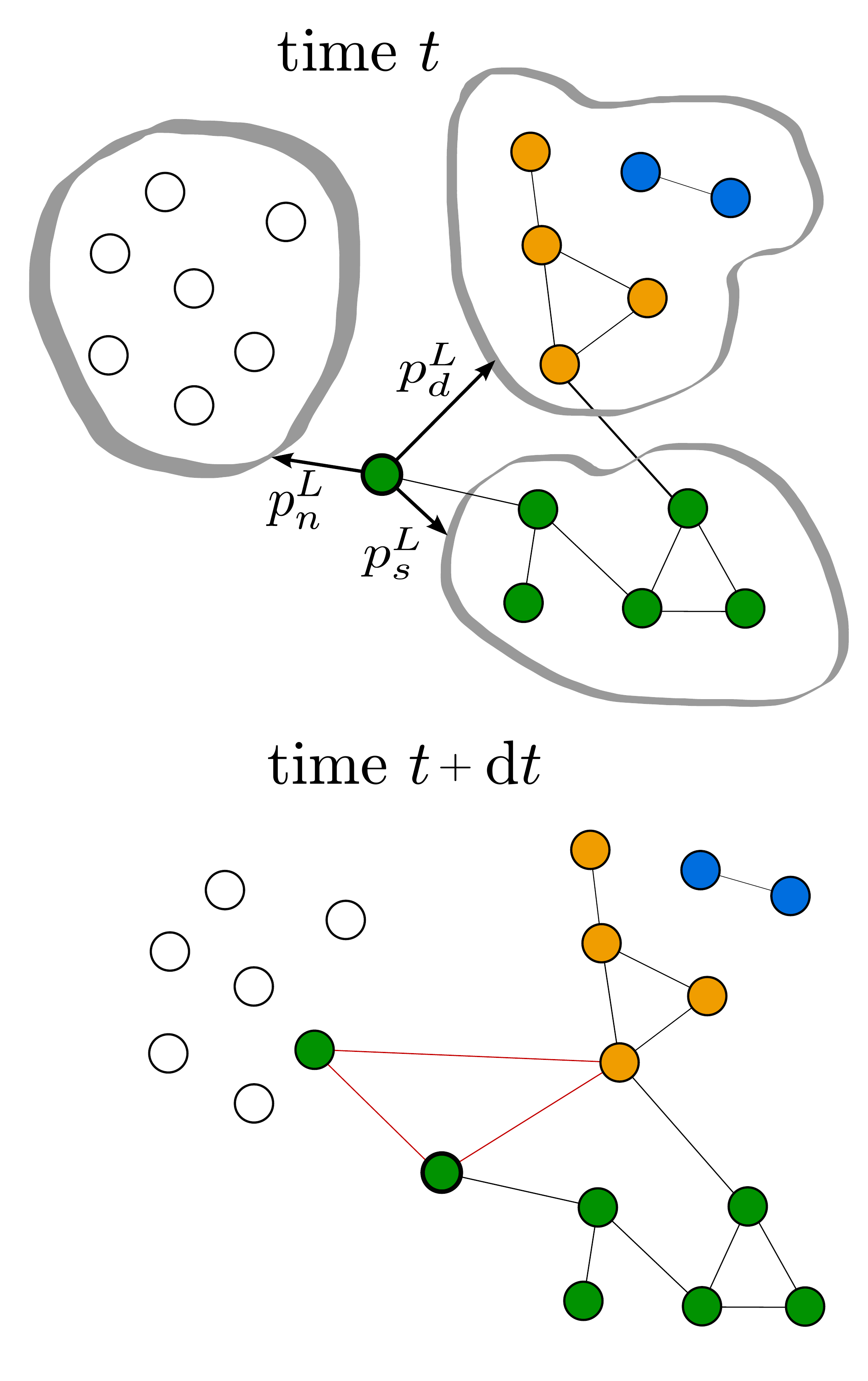}
 \quad\quad
 \footnotesize{(b)}
  \includegraphics[width=0.3\textwidth]{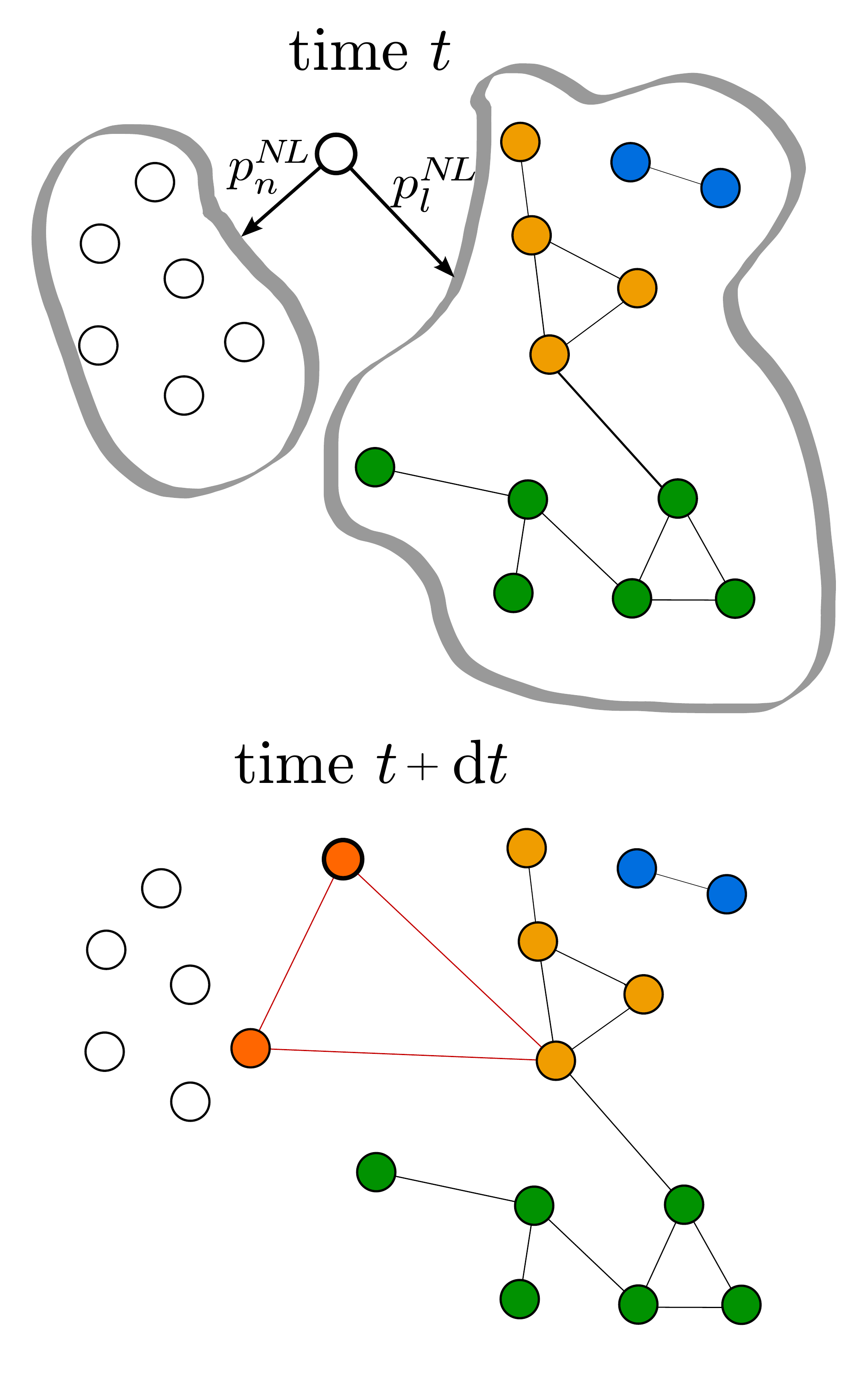}
 \end{center}
 \vspace{-8pt}
 \caption[Two representative examples of collaboration selection and of label propagation.]{Two representative examples of collaboration selection and of label propagation. (a) A labeled agent (whose label is depicted in green) is activated at time $t$ and has to form an alliance with $m=2$ partners. She links to an agent having a different label (depicted in yellow) and one non-labeled, at time $t+\!dt$. (b) Likewise, a non-labeled agent gets activated at time $t\!$ and forms an alliance with $m=2$ partners. She links with one non-labeled agent and one labeled (yellow) agent at time $t\!+\!dt$.}
 \label{fig:model}
 \end{figure}

Figure \ref{fig:model} summarizes the agent-based model described above. 
It illustrates the possible choices for the two different groups, newcomers and established agents.
We note again that this choice progresses in three steps: First, activated agents choose ($m$ times) between newcomers and established agents as partners. 
Subsequently, if activated agents already {have} a label assigned, they have the choice between the group with the same label or groups with a different label. 
{Finally}, within the groups, agents choose their partners with respect to their degree. 
Obviously, the number of agents in each group and the degree of agents change dynamically as the network evolves. 

\section{Model calibration}
\label{sec:model-calibration}

\subsection{Data sources}
\label{sec:data_and_methodology}

{Our agent-based model, as already mentioned, will be calibrated and validated against  data sets from two different domains, covering inter-firm R\&D alliances and co-authorship of scientific papers.}
In the following, we describe the two data sets and afterwards how they are used as input for the model.

\paragraph{R\&D network.}

To reconstruct the R\&D network of collaborating firms we use \textit{SDC Platinum} database.\footnote{\url{http://thomsonreuters.com/sdc-platinum/}} 
It contains data about approximately 672,000 announced alliances from all countries between 1984 and 2009 with daily resolution.
The economic actors participating in these alliances are of several types, e.g. investors, manufacturing firms and universities, but for simplicity we address them as \emph{firms}. 
Each \emph{actor} listed in the data set is associated with a SIC (Standard Industrial Classification) code that allows us to unambiguously assign its corresponding industrial sector.
Further, the purpose of each \emph{alliance} is  characterized by various flags, e.g. manufacturing, licensing, research and development (R\&D).
We restrict ourselves to all alliances with the flag ``R\&D'', which gives us 14,829 alliances connecting 14,561 firms. 
The number of partners involved in each alliance can vary (see Section \ref{sec:input-quantities} for details). 
In most cases the alliance size is two, however it can also be three or higher. 

In order to reconstruct the R\&D network, we focus on the time-aggregated data set. Each firm engaged in a R\&D alliance becomes a node and un-directed links connect nodes involved in the same alliance. 
By adopting this procedure, the 14,829 R\&D alliances result in a total of 21,572 links connecting 14,561 nodes.
To compare collaborations in different industrial sectors, we reconstruct  six distinct R\&D networks for the six largest industrial sectors. 
According to our data set, these are related to computer software, pharmaceuticals, R\&D laboratory and testing, computer hardware, electronic components and communications equipment.
An alliance is considered as part of a given sector if one of the collaborating firms has a matching SIC code. 
The details for the sectoral networks are given in Table \ref{table:collaboration_networks_sizes}. 
Additionally, we compare these sectoral networks with an aggregated R\&D network, previously analyzed by \citet{tomasello2014therole}, which was obtained by considering all the R\&D alliances together, i.e. more than just the six largest industrial sectors. 

\paragraph{Co-authorship network.}

To reconstruct the collaboration network of scientists, we use the data set from the American Physical Society about papers published in any APS journal, namely Physical Review Letters, Reviews of Modern Physics, and all Physical Review journals. (APS).\footnote{\url{http://www.aps.org/}} 
From this data set we use, for each publication, the names of the authors and PACS codes of the papers, to assign the publications to different research areas.
We restrict ourselves to the period from 1983 to 2010, for which we use the time-aggregated data.

This data set has the limitation that the authors are identified by strings which often contain inconsistencies, e.g. missing special characters or spelling mistakes. 
Thus, in order to really make use of the APS data set, we have to  disambiguate authors names in a separate, but time consuming, data processing. 
The latter involves matching the  papers in the APS data set with Microsof Academic Search (MSAS) service, where both papers and authors have unique identifiers. The MSAS is a search engine which mines data from a bibliographic database containing information about scholars and their publications from 15 different disciplines. We have used the Application Programming Interface (API) of MSAS to obtain information about scholars publishing on APS. 
This way, we obtain a list of unique authors that we can use.

To reconstruct the co-authorship network, each unique author is represented by a node and links connect nodes that have co-authored at least one paper in the aggregated data set.  
Following this procedure, the 73,000 papers listed in the data set result in 300,000 links connecting 95,000 nodes.

At difference with the R\&D networks, where firms are characterized by SIC codes, authors are not associated with any classification.
Authors can change their research subject during their career, thus making a categorization on the author level difficult. 
Instead, the classification,  i.e. the PACS number, is assigned to the links of the network representing the papers. 
For this reason, we build co-authorship networks of different fields by using the PACS numbers assigned to papers.
In order to have co-authorship networks comparable in size and density with the R\&D networks, we select the following six representative PACS numbers: 03 (quantum mechanics, field theories and special relativity), 04 (general relativity and gravitation) 42, (optics), 72 (electronic transport in condensed matter), 74 (superconductivity) and 89 (other areas of applied and interdisciplinary physics, that for example includes network theory). We report the sizes of these networks in Table \ref{table:collaboration_networks_sizes}.

\renewcommand{\arraystretch}{1.2}
\begin{table}[htbp]
\centering
\tabcolsep=0.14cm
\begin{tabular}{| l || r|r|r |}
\hline

\textbf{} & $N$ & $E$ & \textit{Links}  \\
\hline
Aggregated R\&D network & 14,561 & 14,829 & 21,572 \\
\hline
\textbf{Sectoral R\&D networks} &  &  &    \\ 
Pharmaceuticals (SIC 283) & 3,829 & 5,277 & 6,019 \\ 
Computer hardware (SIC 357) & 1,582 & 2,672 & 4,047  \\ 
Communications equipment (SIC 366) & 1,133 & 1,888 & 2,726  \\ 
Electronic components (SIC 367) & 1,615 & 2,574 & 3,756 \\ 
Computer software (SIC 737) & 3,381 & 4,134 & 5,862 \\ 
R\&D, laboratory and testing (SIC 873) & 3,188 & 4,032 & 5,364 \\ 
\hline
\textbf{Co-authorship networks} &  &  &    \\ 
Quant. mech., field theories, spec. relativity (PACS 03) & 21,501 & 19,647 & 56,111 \\ 
General relativity and gravitation (PACS 04) & 8,294 & 8,158 & 32,513 \\ 
Optics (PACS 42) & 27,436 & 20,105 & 94,961 \\ 
Electronic transport in condensed matter (PACS 72) & 19,492 & 11,687 & 55,818 \\ 
Superconductivity (PACS 74) & 14,920 & 10,541 & 52,615 \\ 
Other applied and interdisciplin. physics (PACS 89) & 4,881 & 2,873 & 8,777 \\ 
\hline
\end{tabular}
\caption[Number of nodes, collaboration events and links of all examined collaboration networks.]{Number of nodes $N$, of collaboration events $E$ and of resulting links in our representation for the aggregated R\&D network, the six largest sectoral R\&D networks, and the six representative co-authorship networks. For all domains, we consider the respective cumulative networks, i.e. the networks obtained by keeping all the links at any time.}
\label{table:collaboration_networks_sizes}
\end{table}

\subsection{Input quantities}
\label{sec:input-quantities}

Based on the two data sets, we now calculate the two empirical inputs needed for our agent-based model,  namely the 
{size distribution of the collaboration events and the activity distribution of the agents.}

\paragraph{Size of collaboration events.}
In the SDC alliance data set, the size of a collaboration event is the number of firms per R\&D alliance, while in the co-authorship data set it is the number of co-authors per paper.
To study these, we analyzed the distributions of partners per collaboration event, $P(m)$, in both considered data sets.

\begin{figure}[htbp]
\begin{center}
\includegraphics[width=0.8\textwidth]{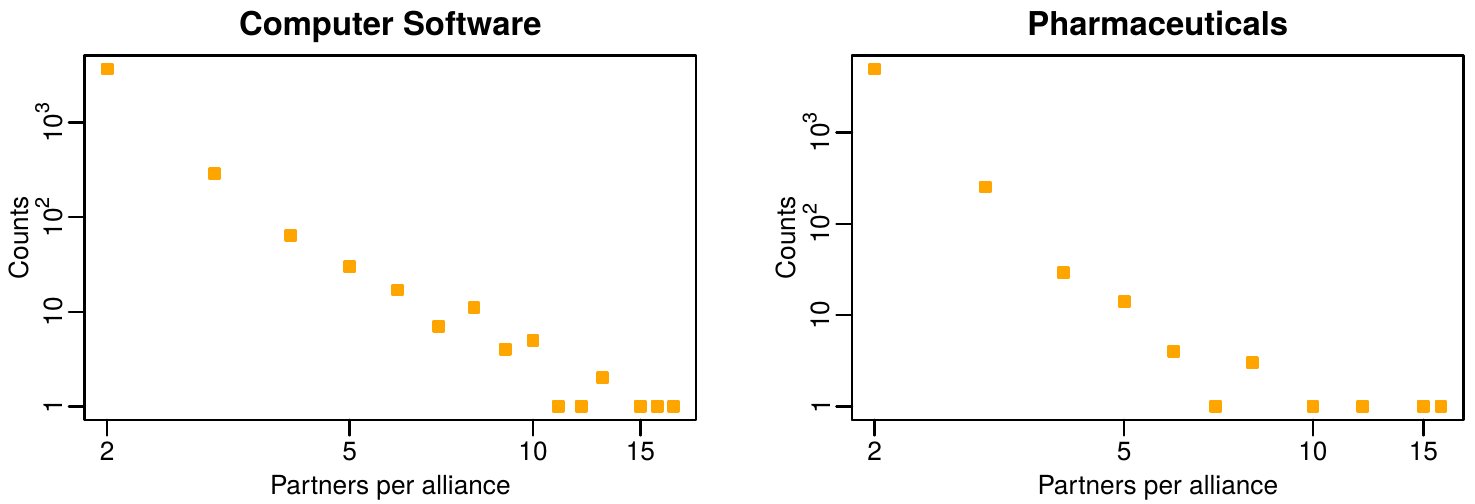} 
\includegraphics[width=0.8\textwidth]{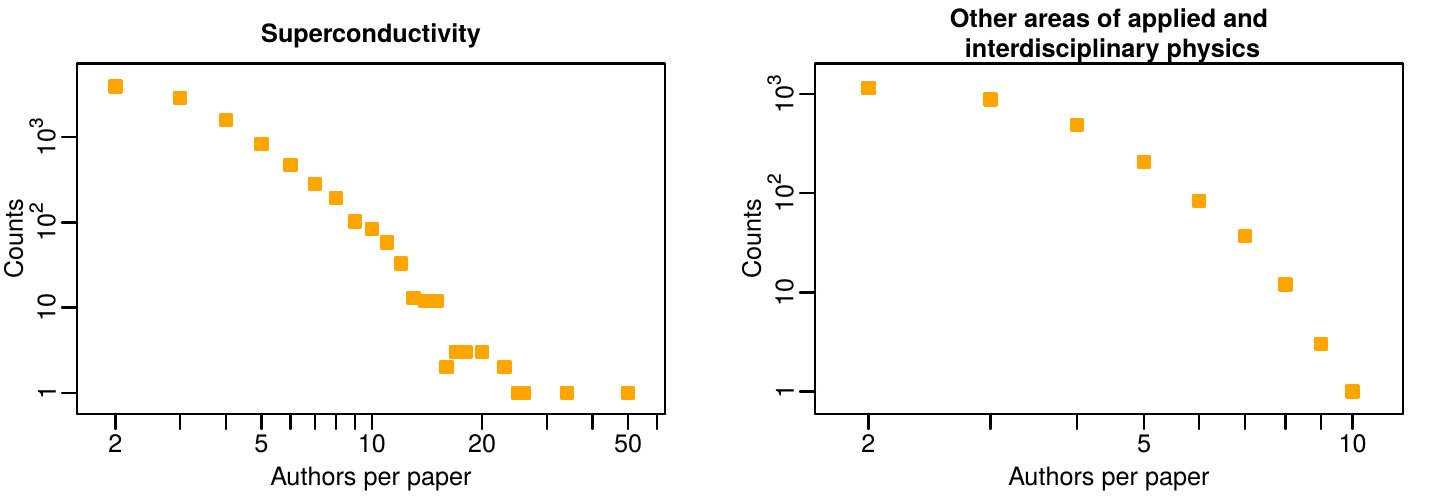} 
\end{center}
\vspace{-16pt}
\caption[Distribution of the number of partners per alliance for two representative industrial sectors and distribution of the number of authors per paper for two representative co-authorship networks.]{Distribution of the number of partners per alliance for two representative industrial sectors: computer software (top left)) and pharmaceutics (top right), as measured from the SDC data set. Distribution of the number of authors per paper for two representative co-authorship networks: superconductivity (bottom left) and interdisciplinary physics (bottom right), as measured from the APS-MSAS data set.}
\label{fig:size_collab_events}
\end{figure}

With respect to our six sectoral R\&D networks, we find that the size distribution is right-skewed with values ranging between 2 and 20. 
It should be noted that the identification of the functional form of these distributions (e.g., power-law, exponential, log-normal and so on) is outside of the scope of this study, therefore we leave it as a possible extension.
Most of the collaborations are stipulated between two partners, but some alliances -- the so-called \textit{consortia} -- involve three or more partners.
In Figure  \ref{fig:size_collab_events} we report such distributions for two represetative industrial sectors. 
Results for four more industrial sectors are presented in Appendix A, confirming that the right-skewed distribution  holds for all sectoral R\&D networks, with only small differences in the tails of the respective distributions.
These results are in line with the ones presented in \citep{tomasello2014therole} for the aggregated R\&D network.

Regarding the size of scientific collaborations, we find results  similar to the R\&D alliances.
I.e., most papers in our APS-MSAS data set have two co-authors with a broad right-skewed size distribution for all PACS numbers investigated.  
From our analysis, we have excluded all  papers written by only one author because we are are interested in collaboration networks, where such papers would only generate isolated nodes. 
For this reason, the counts start from 2 in all of our plots. 
Figure  \ref{fig:size_collab_events} gives representative examples from two PACS numbers. 
Differently from the sectoral R\&D networks, the co-authorship networks exhibit a larger degree of variability among PACS numbers. 
This is due to the fact that the typical number of authors per paper strongly depends on the field. 
To give an example, the field of applied and interdisciplinary physics is characterized by significantly fewer authors per paper (at most 10) than the field of general relativity and gravitation (whose right tail reaches 55 authors per paper).
In Figure  \ref{fig:sectoral_partners} and Figure  \ref{fig:authors_per_paper_distr} in Appendix A, we show the distribution of collaboration sizes for respectively the
six sectoral R\&D networks and the six co-authorship networks.

\paragraph{Agents' activity.}
This is one of the two key attributes assigned to agents in our model. 
We apply a measure developed in the setting of temporal networks \citep{holme2012temporal}, which has been already used to analyze various data sets \citep{barabasi2005origin, barabasi99:_emerg, pastor-satorras01:_dynam_correl_proper_inter}, also in  the context of R\&D and co-authorship networks \citep{tomasello2014therole, perra2012activity}.

\begin{figure}[htbp]
\begin{center}
\includegraphics[width=0.80\textwidth]{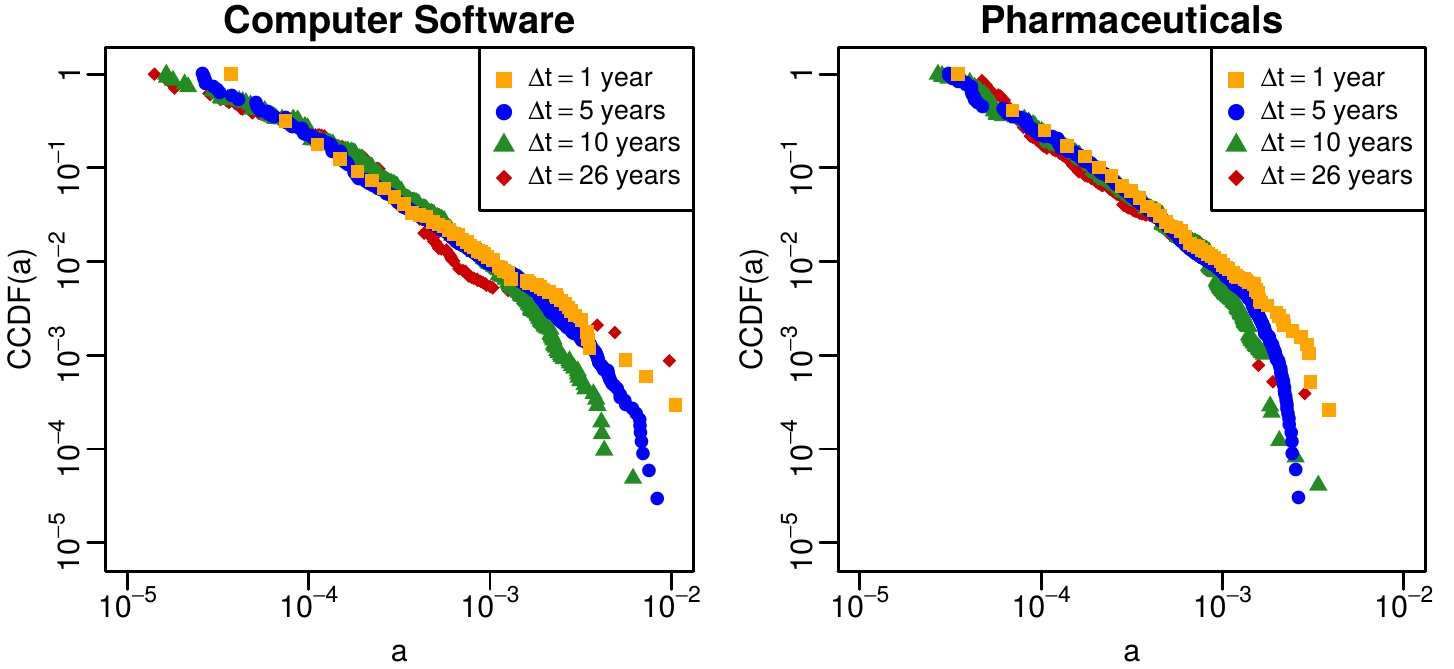}
\includegraphics[width=0.80\textwidth]{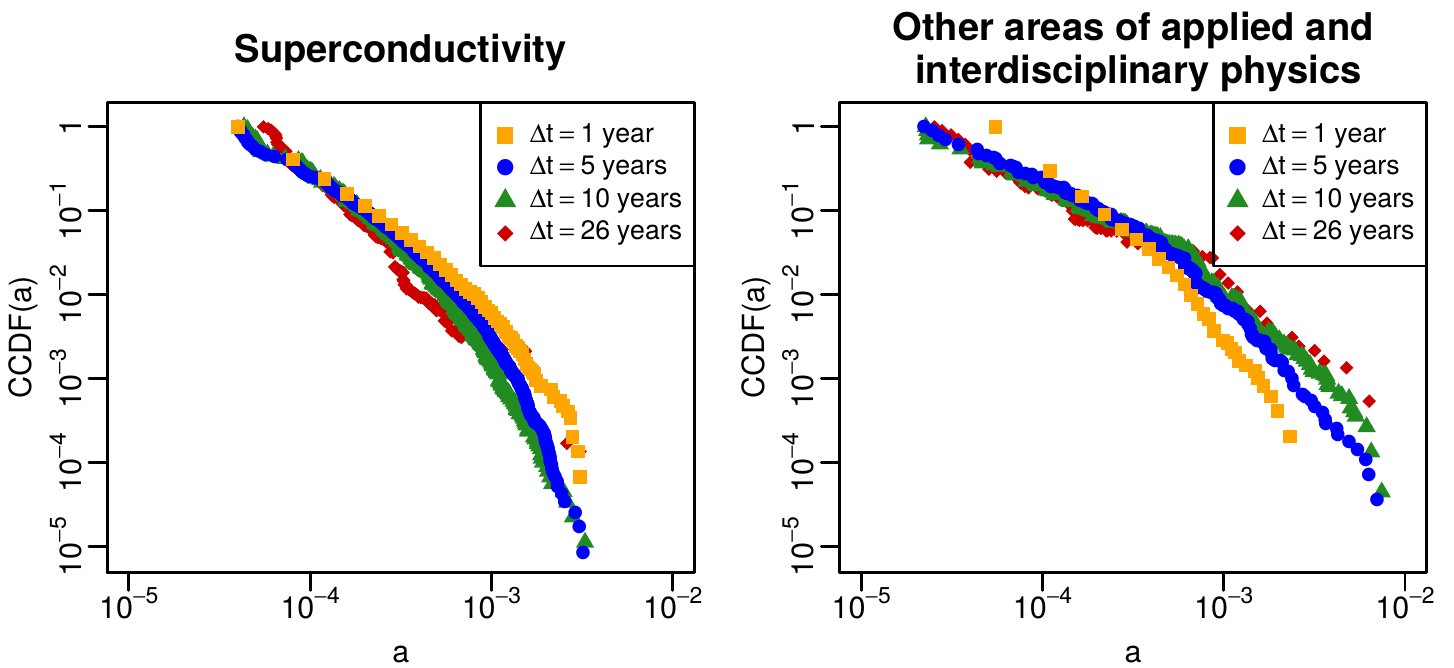}
\end{center}
\vspace{-16pt}
\caption[Agent activity distribution in the two representative sectoral R\&D networks.]{Complementary cumulative distribution function (CCDF) of the empirical firm activities, measured for two representative industrial sectors (from the SDC data set, \citep{tomasello2014therole} Supplementary information), and of the empirical author activities, measured for two representative co-authorship networks (from the APS-MSAS data sets). 
We considered for 4 different time windows $\Delta t$ of 1, 5, 10 and 26 years. 
}
\label{fig:activities_rep}
\end{figure}

Following these approaches, we argue that  activity reflects the propensity of an agent to participate in a collaboration event. 
Precisely, we define the empirical activity of an agent $i$ at time $t$ as the number of collaboration events, $e^{\Delta t}_{i,t}$, involving agent $i$ during a time window $\Delta t$ ending at time $t$ divided by the \emph{total number} of collaboration events, $E^{\Delta t}_{t}$, involving any agent during the same period of time:
\begin{equation}
 a^{\Delta t}_{i,t} =  \frac{{e^{\Delta t}_{i,t}}}{ {E^{\Delta t}_{t}}}.
\end{equation}
For both the SDC alliance and APS-MSAS data sets, we measure the empirical distribution of activity, $P(a)$, for four different time windows, $\Delta t=1,5,10$ and 26. 
When the time window is shorter than 26 years (the entire data set observation period), we compute the activity by shifting the time window in 1-year increments and then we average the results.
For simplicity, from now on, we will write $a^{\Delta t=26 years}_{i,2009}$ as $a_i$, which is the activity over the longest time window.
Interestingly, we find that these distributions are independent of the size of the time window, which is a robust feature for both R\&D and co-authorship collaborations. 
In Figure \ref{fig:activities_rep}, we report these results for two representative sectoral R\&D networks and two representative co-authorship networks.
For a visualization of the complete results for the six sectoral R\&D networks see \citep{tomasello2014therole} (Supplementary information) and for the six co-authorship networks see Figure  \ref{fig:coauthorship_activities} in Appendix A.

\subsection{Implementation and optimal model selection}
\label{sec:impl-choos-optim}

To reproduce the collaboration networks from the two domains, we implement the agent-based model described in Section \ref{sec:agent-based-model}. 
For the simulations, we take the number of agents, $N$, and the total number of collaboration events, $E$, from the respective empirical networks. 
The two input parameters, size of the collaboration event, $m_{i}$, and agent activity, $a_{i}$, are obtained by sampling from the above distributions, $P(m)$ and $P(a)$.  
With that, 
the only free parameters in our model are the five probabilities $p_{s}^{L}$, $p_{d}^{L}$, $p_{n}^{L}$, $p_{n}^{NL}$, $p_{n}^{NL}$ which we vary in order to find which combination gives the best match between the simulated and the observed network. 
For more information about the exploration of the parameter space see Appendix B.
For the comparison we use the following quantities: average degree, $\mean{k}$, average path length, $\mean{l}$, and global clustering coefficient, $C$, and define the respective relative errors $\epsilon_{\left\langle k \right\rangle}$, $\epsilon _{\left\langle l \right\rangle}$ and $\epsilon_C $ between the observed and the simulated quantities. 
We require that these errors have to be smaller than a threshold $\epsilon ^0$. 
For all probability combinations we perform  25 simulations. 
We then select the combination that gives us the highest fraction of networks that match the criterion $\epsilon<\epsilon_{0}$. 
The optimal probabilities are indicated using a {star (e.g. $p*_s^L$)}.

\begin{table}[htbp]
\centering
\tabcolsep=0.1cm
\begin{tabular}{|l || r|r|r || r|r| }
\hline
\textbf{} & \boldmath$\quad p^{*L}_{s}$ & \boldmath$p^{*L}_{d}$ & \boldmath$p^{*L}_{n}$ & \boldmath$p^{*N\!L}_{l}$ & \boldmath$p^{*N\!L}_{nl}$ \\
\hline
Aggregated R\&D network & 0.30 & 0.30 & 0.40 & 0.75 & 0.25 \\
\hline
\textbf{Sectoral R\&D networks} & & & & & \\ 
Pharmaceuticals (SIC 283) & 0.35 & 0.35 & 0.30 & 0.80 & 0.20 \\ 
Computer hardware (SIC 357) & 0.55 & 0.30 & 0.15 & 0.90 & 0.10 \\ 
Communications equipment (SIC 366)  & 0.75 & 0.15 & 0.10 & 0.80 & 0.20 \\ 
Electronic components (SIC 367) & 0.65 & 0.20 & 0.15 & 0.90 & 0.10 \\ 
Computer software (SIC 737) & 0.55 & 0.20 & 0.25 & 0.95 & 0.05 \\ 
R\&D, laboratory and testing (SIC 873) & 0.40 & 0.40 & 0.20 & 0.20 & 0.80 \\ 
\hline
\textbf{Co-authorship networks} & & & & & \\ 
Quant. mech., field theor., spec. relativity (PACS 03) & 0.85 & 0.05 & 0.10 & 0.45 & 0.55 \\ 
General relativity and gravitation (PACS 04)$^\dagger{}$  & \textit{0.50} & \textit{0.05} & \textit{0.45} & \textit{0.05} & \textit{0.95} \\ 
Optics (PACS 42) & 0.60 & 0.05 & 0.35 & 0.35 & 0.65 \\ 
Electronic transport in condensed matter (PACS 72) & 0.50 & 0.05 & 0.45 & 0.30 & 0.70 \\ 
Superconductivity (PACS 74)  & 0.55 & 0.05 & 0.40 & 0.35 & 0.65 \\ 
Other applied and interdisciplin. physics (PACS 89) & 0.65 & 0.05 & 0.30 & 0.25 & 0.75 \\ 
\hline
\end{tabular}
\caption[Summary of all optimal simulated network statistics.]{Optimal sets of probabilities to simulated the collaboration networks. The optimal probabilities are indicated using $^*$. 
Recall that the probability of a labeled agent to select an agent with the same label is $p^{L}_{s}$, to select an agent with a different label is $p^{L}_{d}$) and to select a non-labeled agent is $p^{L}_{n}$). While, the probability of a non-labeled agent to select a labeled agent is $p^{N\!L}_{l}$ and to select a non-labeled agent is $p^{N\!L}_{nl}$.
The probabilities $p^{L}_{s}$, $p^{L}_{d}$ and $p^{L}_{n}$ sum up to 1; likewise, $p^{N\!L}_{l}$ and $p^{N\!L}_{nl}$ sum up to 1.\\
$^\dagger{}$ {\scriptsize Only for the co-authorship network in general relativity and gravitation (PACS 04) 
the model is unable to generate a network matching all the three measures $\mean{k}$, $\mean{l}$ and $C$ at the same time. Only $\mean{l}$ and $C$ can be retrieved with an accuracy of 30\%, while the generated $\mean{k}$ is not compatible with the empirical measure.  Even though we report these values for completeness, they cannot be considered significant.

}
}
\label{table:collaboration_networks_optimal_parameters}
\end{table}

In Table \ref{table:collaboration_networks_optimal_parameters} we report the optimal set of probabilities for the collaboration networks from the two different domains. 
The network simulated using the optimal set of probabilities will be named \textit{optimal simulated networks}.
In Table \ref{table:collaboration_networks_empirical} in Appendix B, we report the $\mean{k}$, $\mean{l}$ and $C$ of the optimal simulated networks and they can be compared with the respective values for the observed networks.  
With this, we are set for the \emph{validation} of our agent-based model which of course has to include features of the network that were not used for the \emph{calibration} of the model.

\section{Model validation}
\label{sec:model-validation}
\subsection{Reproducing four distributions}

To validate our agent-based model, we compare the empirical networks with the statistical properties of the simulated ones using the optimal set of probabilities. 
For the comparison, we use macroscopic features such as distributions of degrees, path lengths, local clustering coefficients and sizes of the disconnected components. 
Additionally, we also investigate microscopic, or agent centric, features such as labels. 
The validation procedure is similar to the one described in \citep{tomasello2014therole}.
{To validate the above mentioned distributions,} we emphasize that for the calibration we did \emph{not} use information about the distributions, but only about the respective average values, $\mean{k}$, $\mean{l}$ and $C$, to calculate the relative errors. 

\begin{figure}[htbp]
\begin{center}
\footnotesize{
(a) \includegraphics[width=0.35\textwidth,angle=0]{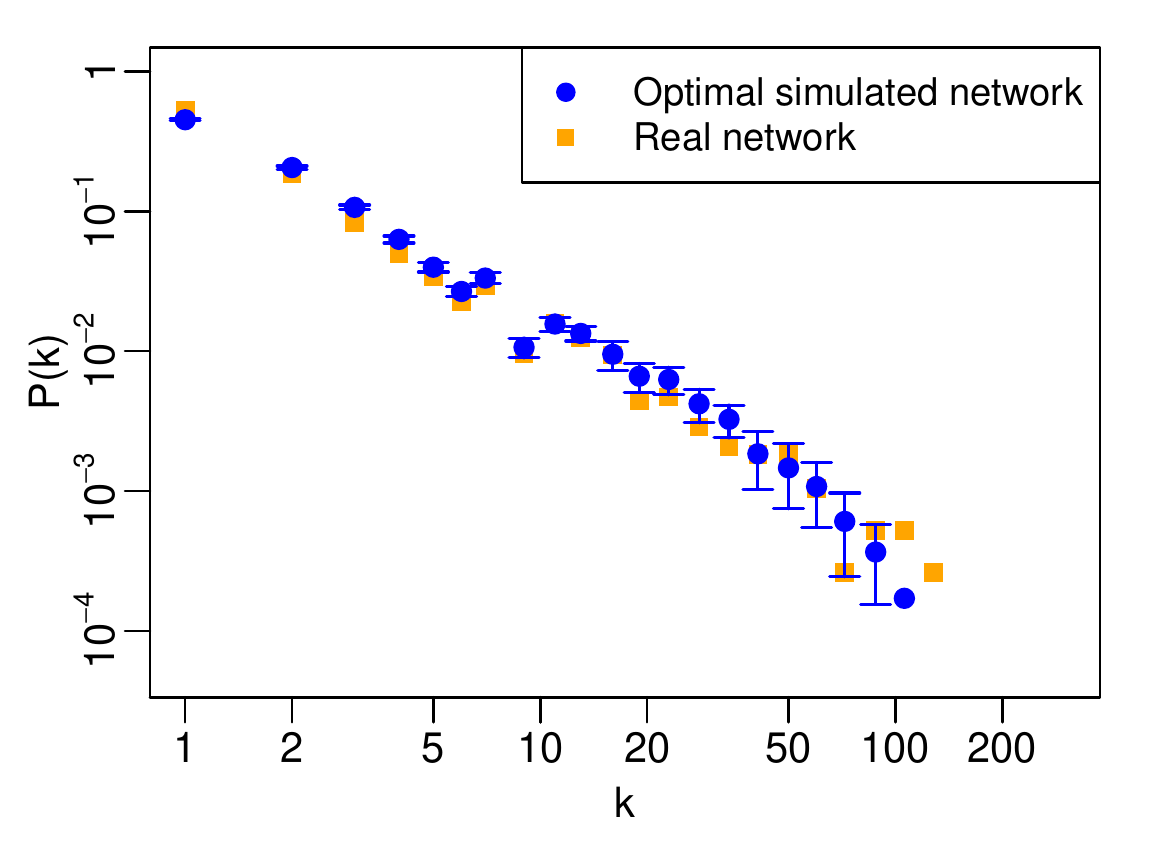}
(b) \includegraphics[width=0.35\textwidth,angle=0]{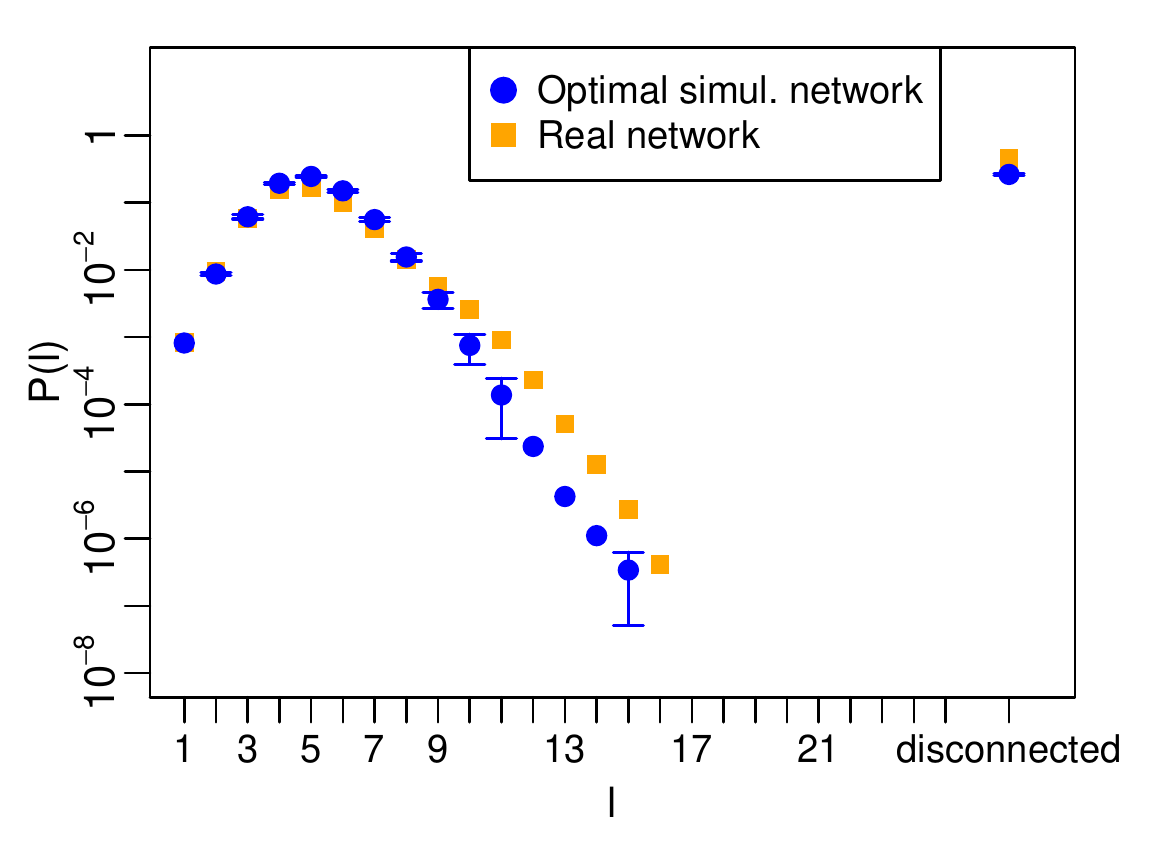}\\
(c) \includegraphics[width=0.35\textwidth,angle=0]{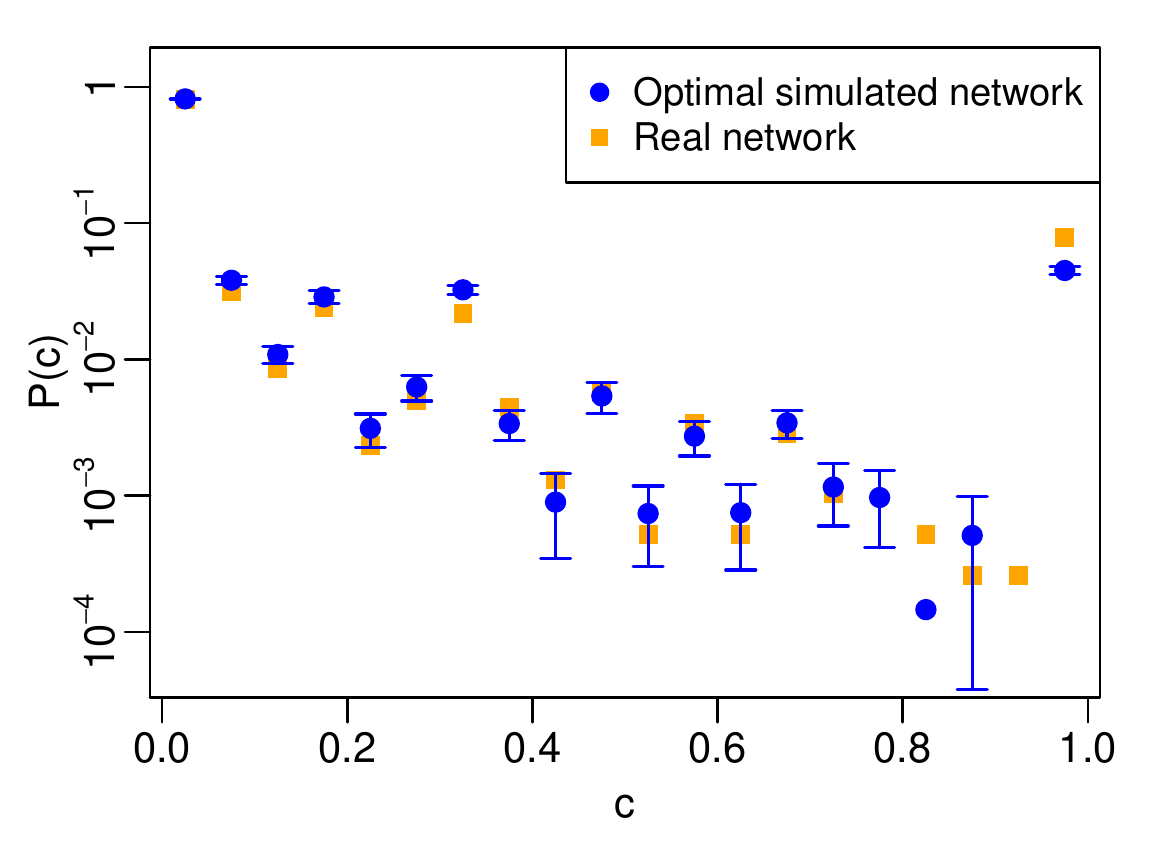}
(d) \includegraphics[width=0.35\textwidth,angle=0]{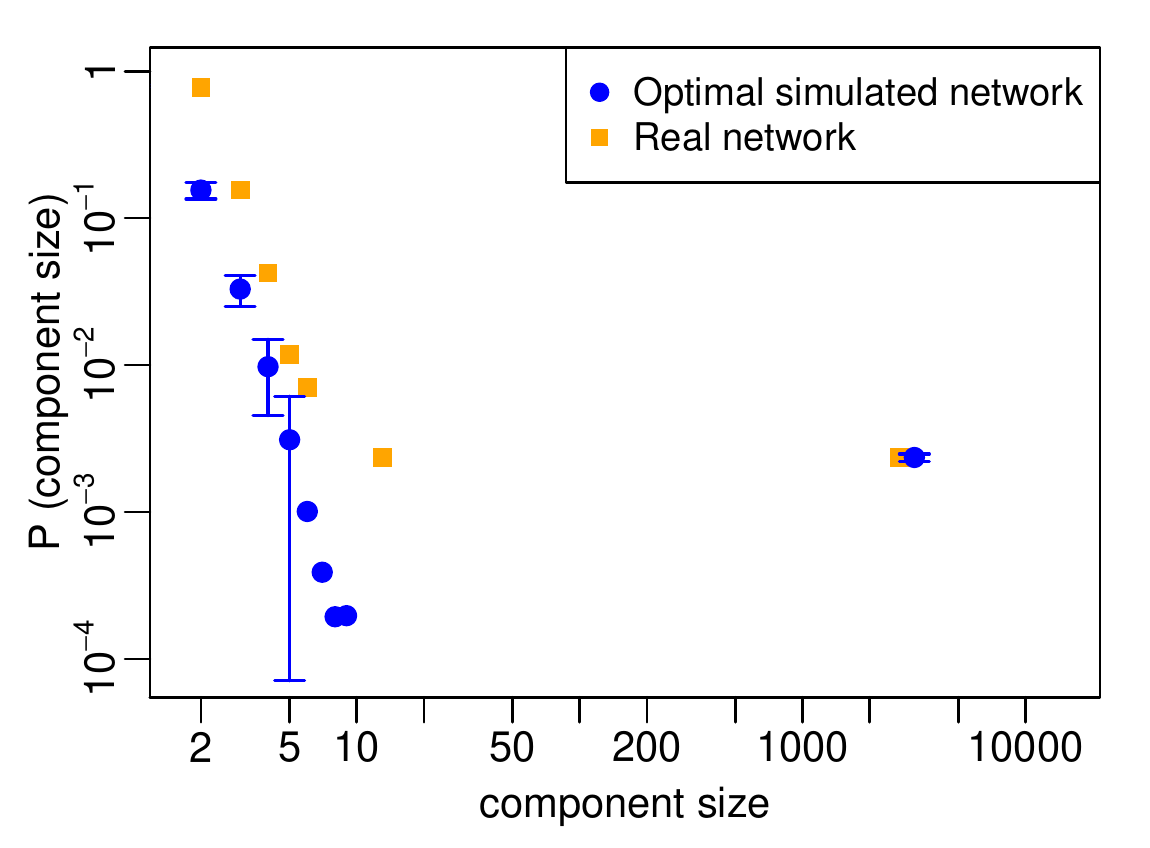}
}
\end{center}
\vspace{-8pt}
\caption[Macroscopic measures for the Pharmaceuticals sectoral R\&D network.]{Distributions of node degrees (a), path lengths (b), local clustering coefficients (c) and component sizes (d) for the real and the 25 optimal simulated networks in ``Pharmaceuticals''(SIC code 283). The blue circles in our plots correspond to the mean values and the error bars correspond to the standard deviations of all the quantities we analyze on the 25 realizations of each optimal simulated collaboration network. In many cases, the error bars are not visible, because the values are very narrowly distributed across these 25 realizations.}
\label{fig:sect_net_pharma}
\end{figure}

\begin{figure}[htbp]
\begin{center}
\footnotesize{
(a) \includegraphics[width=0.35\textwidth,angle=0]{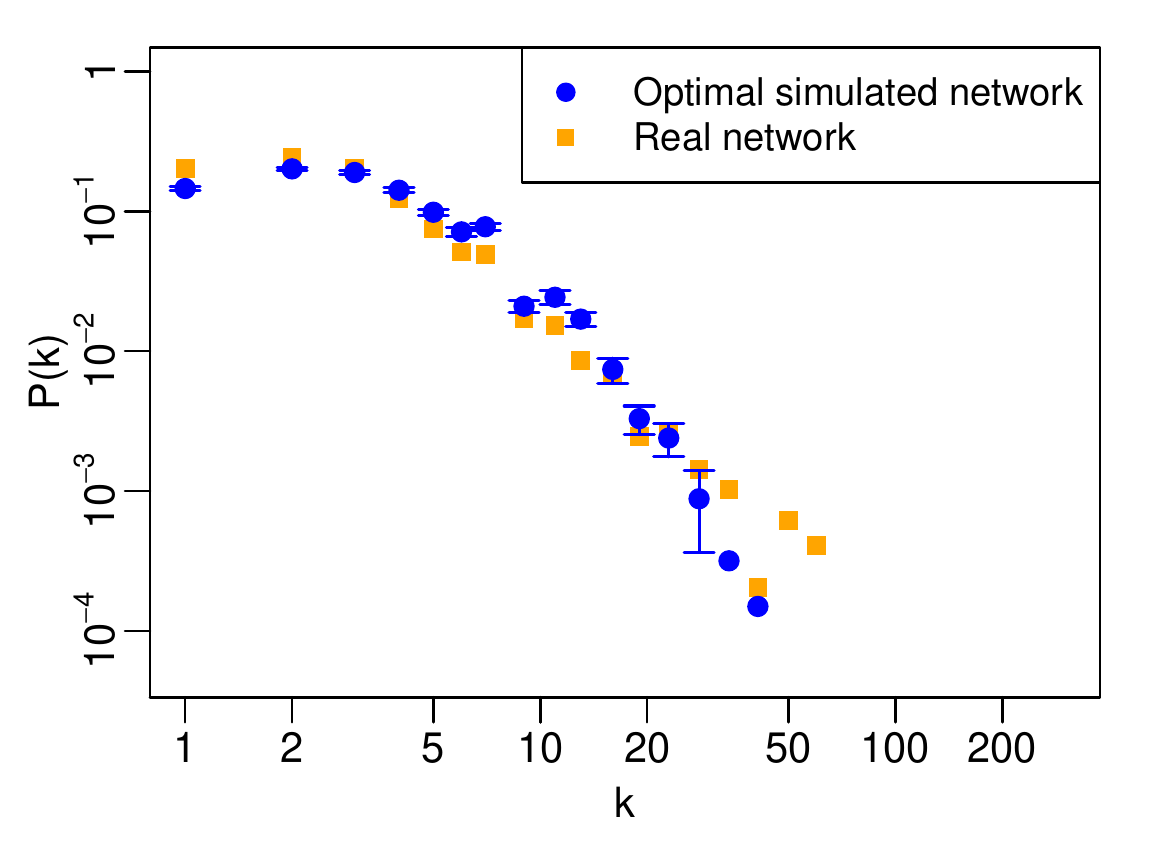}
(b) \includegraphics[width=0.35\textwidth,angle=0]{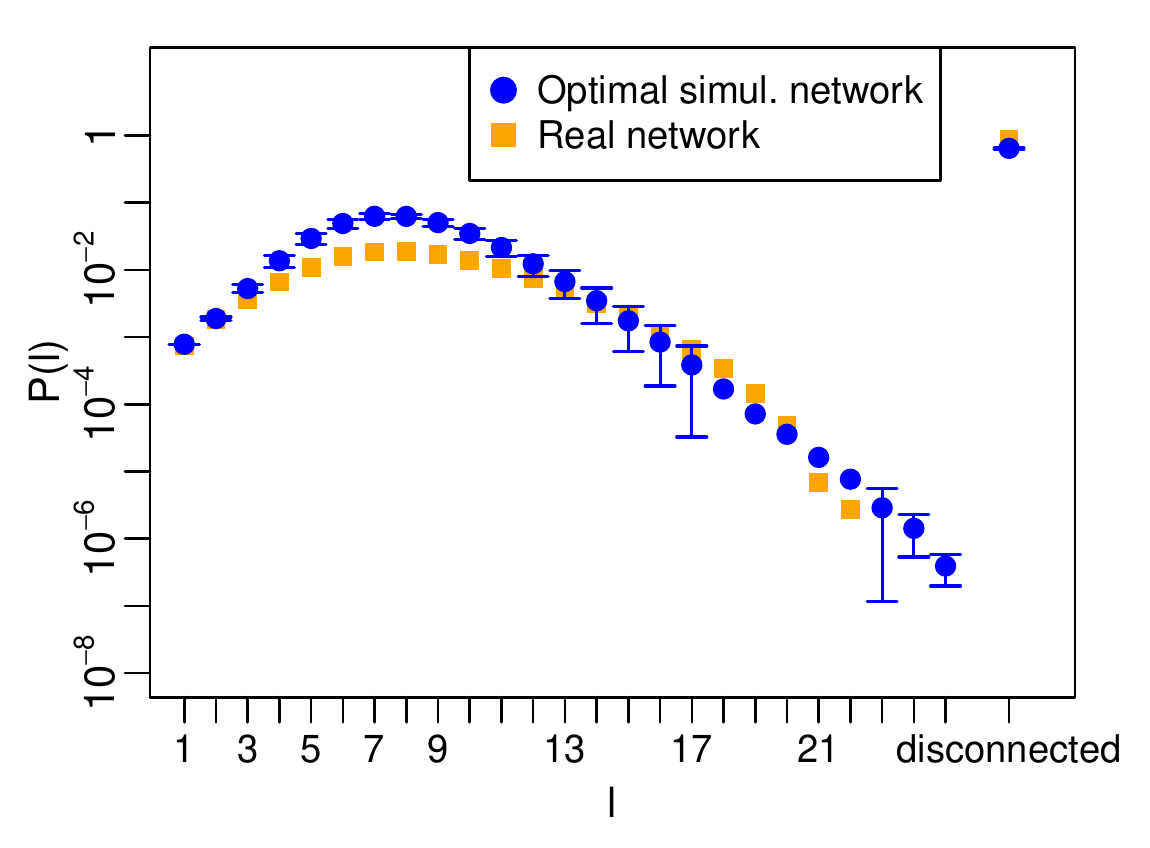}\\
(c) \includegraphics[width=0.35\textwidth,angle=0]{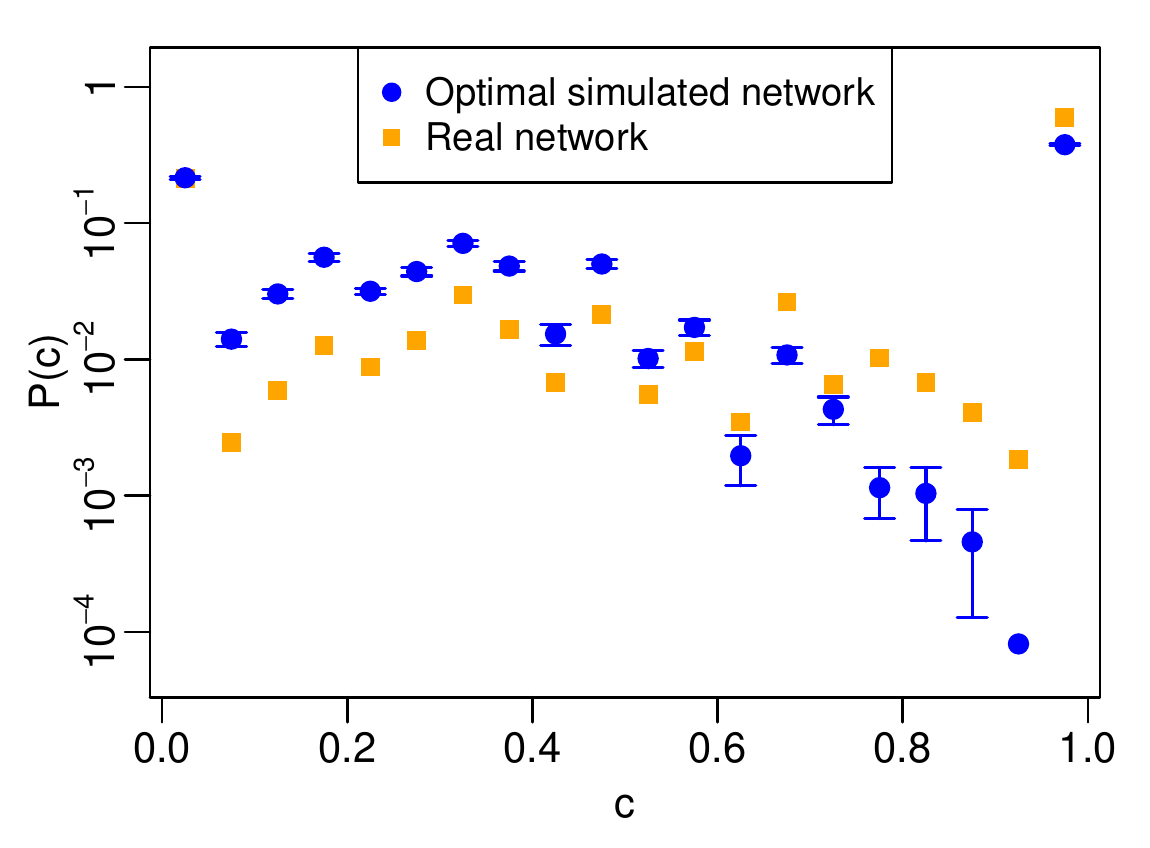}
(d) \includegraphics[width=0.35\textwidth,angle=0]{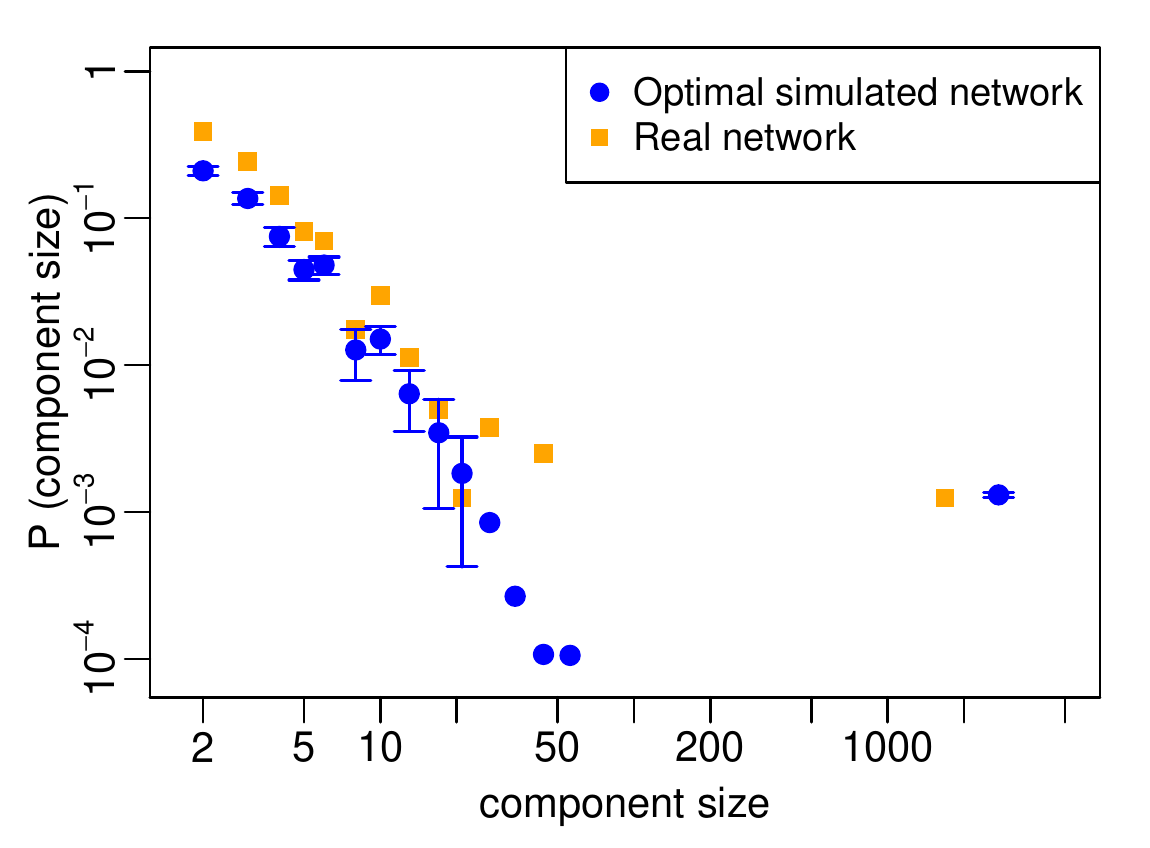}
}
\end{center}
\vspace{-8pt}
\caption[Macroscopic measures for the applied and interdisciplinary physics co-authorship network.]{Distributions of node degrees (a), path lengths (b), local clustering coefficients (c) and component sizes (d) for the real and the 25 optimal simulated networks in applied and interdisciplinary physics (PACS number 89).}
\label{fig:sect_net_PACS89}
\end{figure}

Figure  \ref{fig:sect_net_pharma} and Figure  \ref{fig:sect_net_PACS89} show these distributions for one representative sectoral R\&D network and one co-authorship network.
We observe a remarkable match between the simulated and the empirical distributions for all four quantities. 
In particular, the model reproduces the emergence of a giant component in both networks, together with many smaller components down to size two. 

\subsection{Community structures and groups of influence}
\label{sec:comm-str-and-groups-of-infl}
The second part of our validation regards the modular structure of the collaboration networks in terms of communities. 
We start by evaluating and comparing the community structure of the observed networks and of the simulated ones using the optimal set of probabilities.
Then, we verify that the groups of influence defined by the agents' labels well reproduce the community structure of the simulated networks.

\begin{figure}[htbp]
\begin{center}
(a) \includegraphics[width=0.45\textwidth,angle=0]{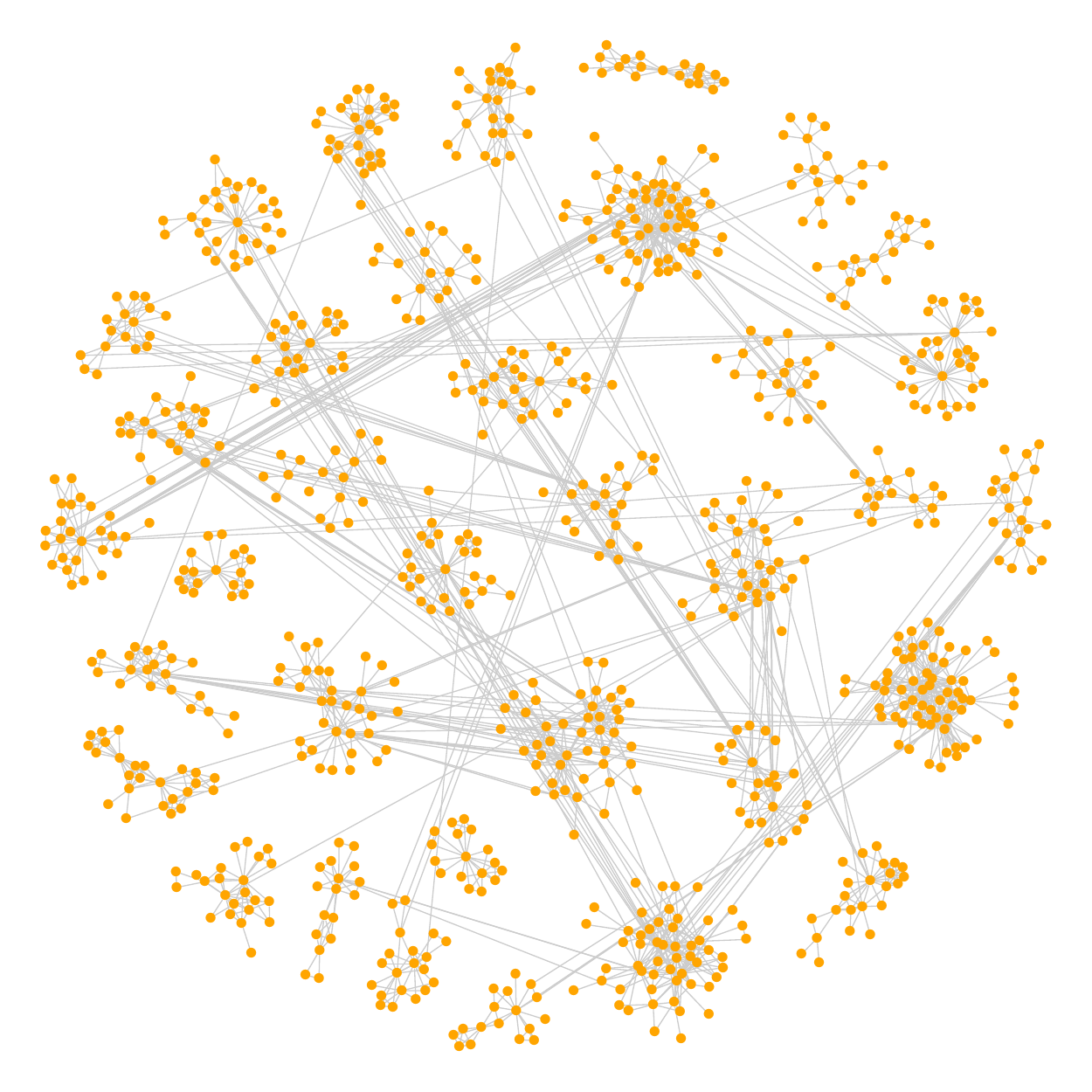}
(b) \includegraphics[width=0.45\textwidth,angle=0]{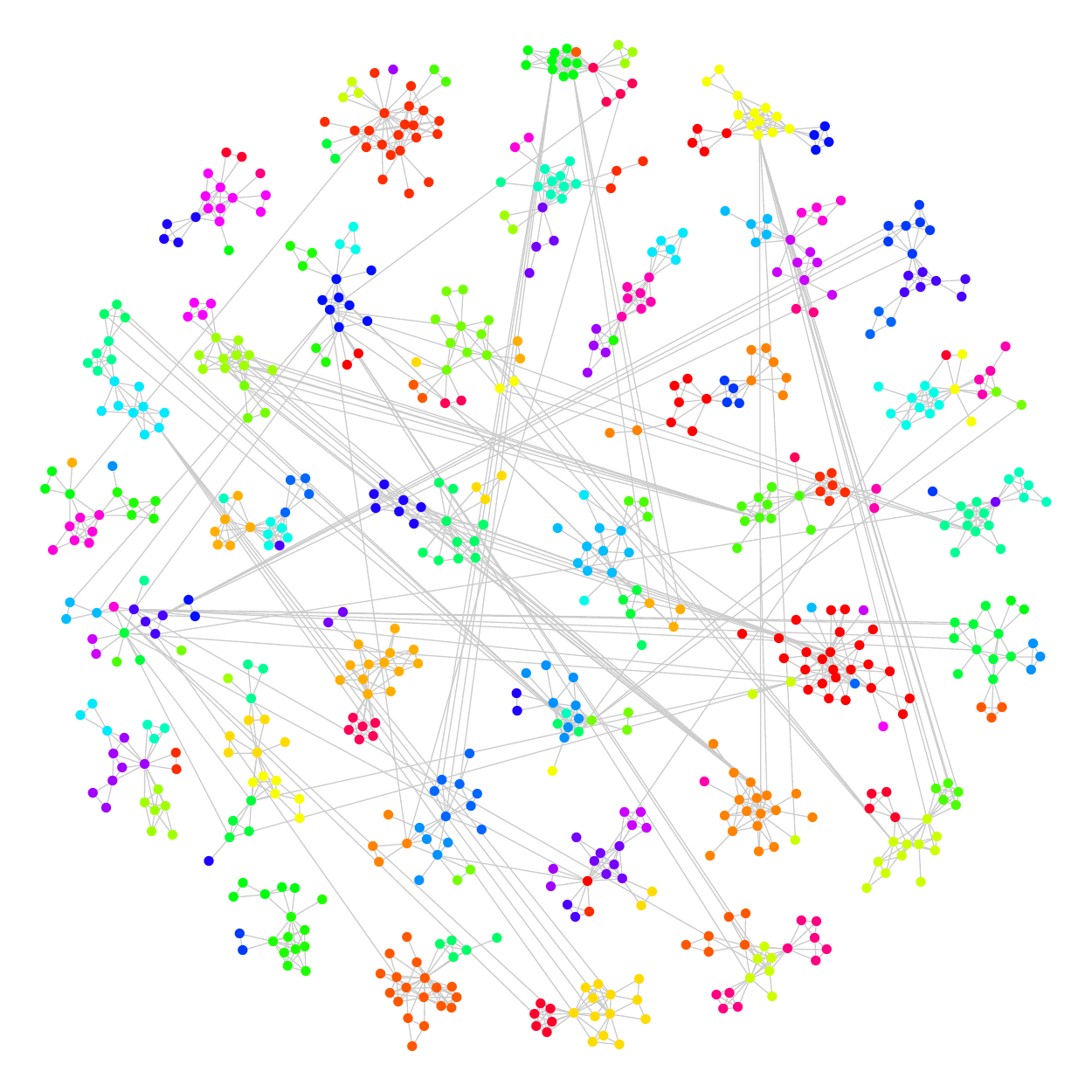}
\end{center}
\vspace{-8pt}
\caption[Network visualization and communities in the co-authorship network of applied and interdisciplinary physics: empirical and simulated.]{Co-authorship network in applied and interdisciplinary physics (PACS number 89).(a) Visual representation of the empirical network, considering only the 30 largest clusters detected by the \texttt{Infomap} algorithm. Distinct clusters are represented by grouping nodes in distinct regions of the plot area. (b) Visual representation of one realization of the simulated network, considering only the 30 largest clusters detected by the \texttt{Infomap} algorithm. Distinct clusters are represented by node groups in distinct regions of the plot area. In addition, we depict our node labels by using different colors: most of the nodes in a given cluster share the same label.}
\label{fig:COAUTnet89_empirical_simulated_communities}
\end{figure}

\paragraph{Community structure of empirical and simulated networks.} 
To detect the community structure in the observed networks,  we employ a widely used algorithm, \texttt{Infomap} \citep{rosvall2008infomap}, which is based on the probability flow of random walks on networks.
In Table \ref{table:collaboration_networks_clusters} in Appendix C, we report the number of communities found in each network. 
In Figure  \ref{fig:COAUTnet89_empirical_simulated_communities} (a), we give a visual representation 
of the respective communities in the co-authorship network in applied and interdisciplinary physics.

In order to quantify the goodness of the community partitions detected by \texttt{Infomap}, we use a normalized modularity score $Q$.
This coefficient is equal to 1 when all links connect only nodes belonging to the same community, equal to 0 for a network where links are placed randomly, and equal to -1 when links are formed only among nodes populating distinct communities.
Interestingly, we find that \emph{all} the R\&D \emph{and} co-authorship networks are characterized by a high modularity as reported in Table \ref{table:collaboration_networks_clusters} in Appendix C.
Precisely, all the $Q$ scores for partitions originated by \texttt{Infomap} are significantly higher than the equivalent scores on randomly generated networks with the same degree sequence, especially in the domain of co-authorship networks.
We can safely conclude that our high $Q$ values are indicative of a real modular structure, and not a simple artifact of the network's size and density\citep{reichardt2006networks}.

\begin{figure}[htbp]
\begin{center}
\footnotesize{
(a) \includegraphics[width=0.45\textwidth,angle=0]{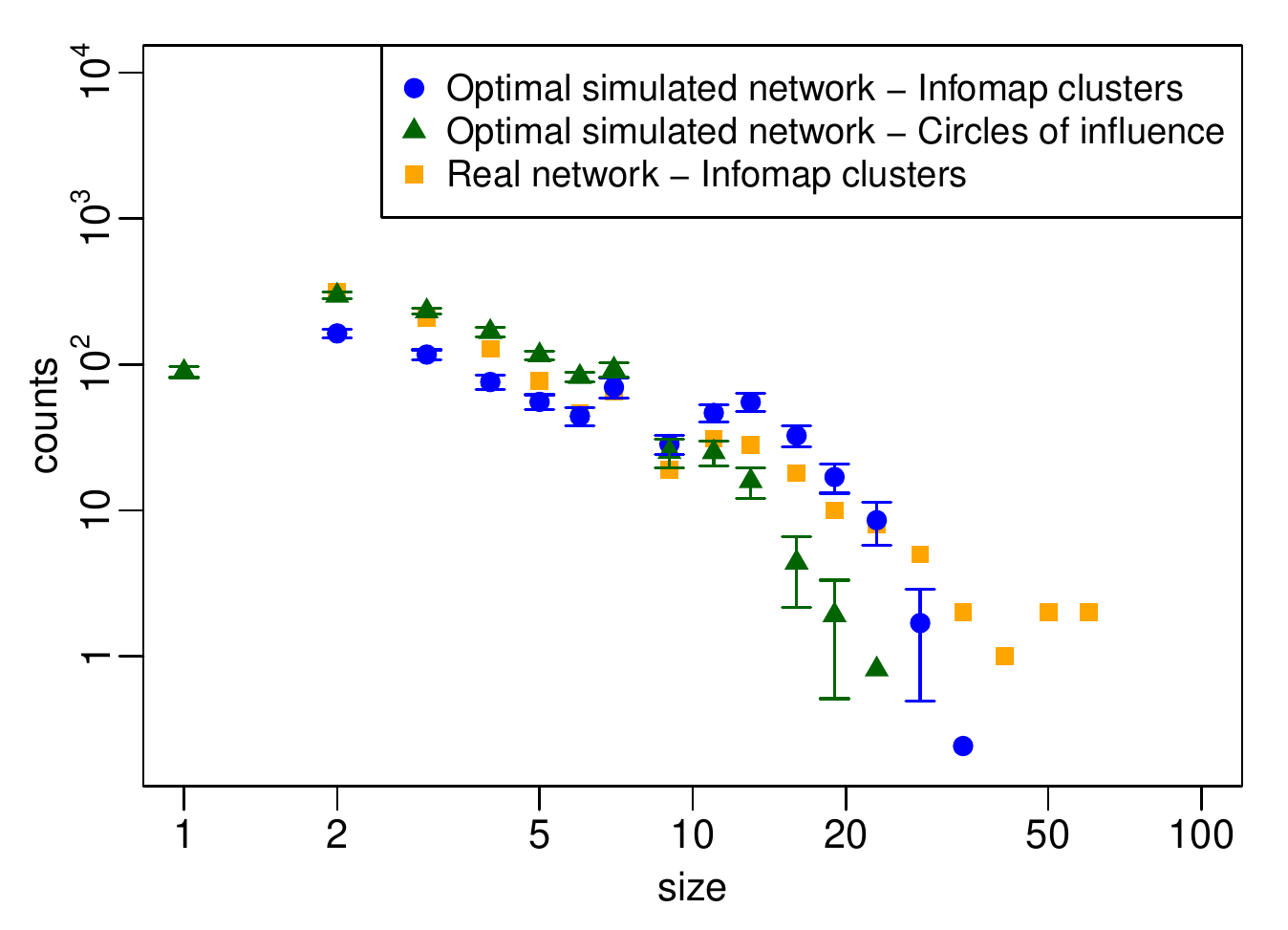}
(b) \includegraphics[width=0.45\textwidth,angle=0]{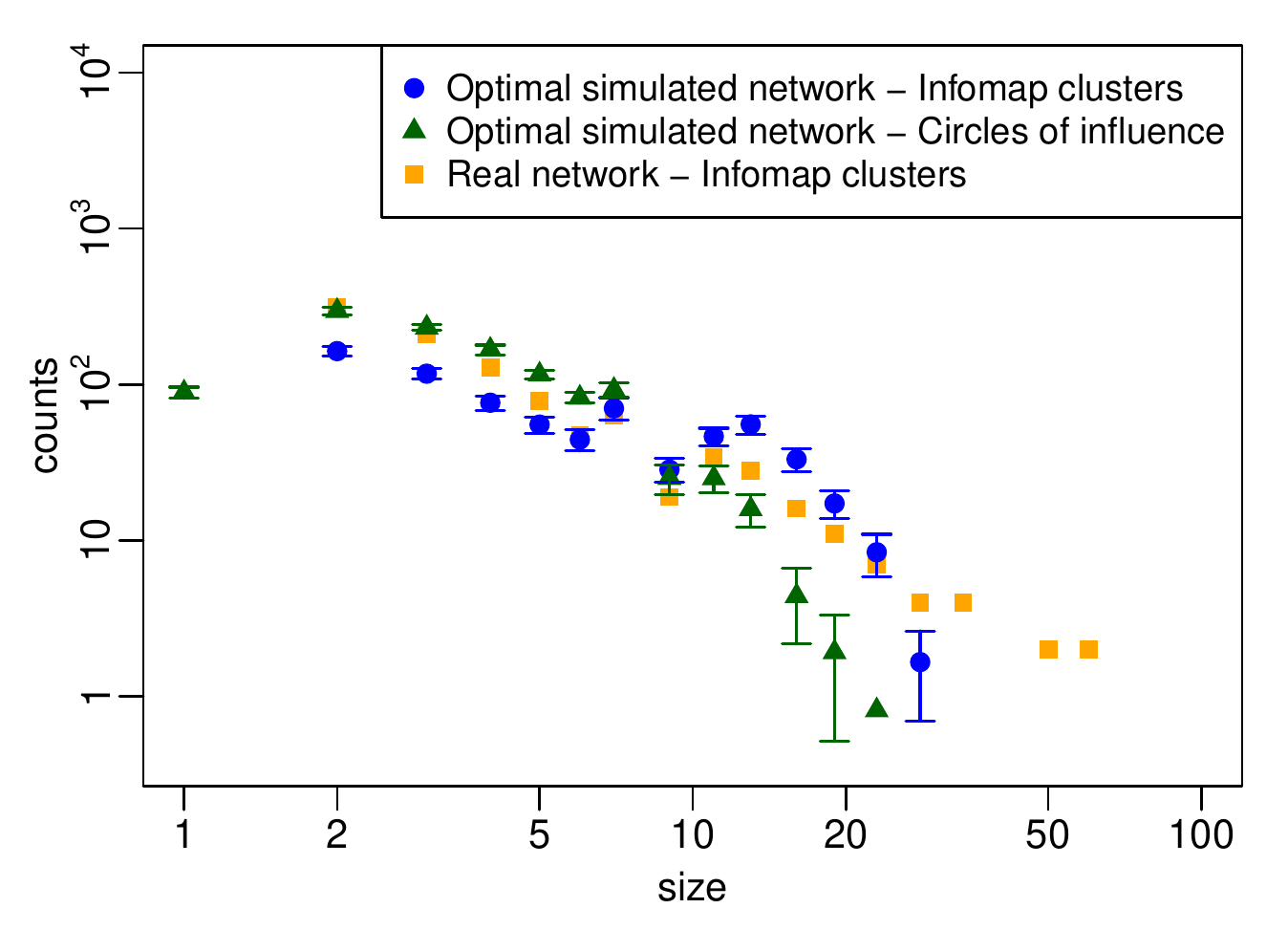}
}
\end{center}
\vspace{-8pt}
\caption[Distributions of communities size for the ``Pharmaceuticals '' R\&D network and for the co-authorship network in applied and interdisciplinary physics.]{Size distribution of i. the circles of influence in the 25 realizations of the optimal simulated network, ii. the \texttt{Infomap} clusters in the 25 realizations of the optimal simulated network and iii. the \texttt{Infomap} clusters in the empirical network for ``Pharmaceuticals'' (SIC code 283) (a) and ``Applied and interdisciplinary physics'' (PACS number 89) (b).}
\label{fig:distributions_of_communities}
\end{figure}

To detect communities structure on the \emph{simulated} networks, we employ the same procedure we have described above.
We visualize the partitioning detected for the co-authorship network in other applied and interdisciplinary physics in Figure  \ref{fig:COAUTnet89_empirical_simulated_communities} (b). 
The simulated distributions of clusters size match their empirical counterparts, which is far from being trivial given that no information about the community structure was used for the calibration. We report this result for the ``Pharmaceuticals'' R\&D network in Figure  \ref{fig:distributions_of_communities} (a), and for the co-authorship network in applied and interdisciplinary physics in Figure  \ref{fig:distributions_of_communities} (b).

Another evidence of their similarity is the modularity score of the optimal simulated networks -- $Q^{*}=0.61 \pm 0.01$ for the Pharmaceuticals R\&D network, and  $Q^{*}=0.87 \pm 0.01$ for the co-authorship network in interdisciplinary physics.
These values are close to their empirical equivalents, 0.62 and 0.92 respectively. 
In all cases, the modularity scores are significantly greater (with a $p$-value computationally indistinguishable from zero) than the ones obtained for a set of 100 randomly generated networks with the same degree sequence, proving that the obtained modularity cannot be expected or explained simply with the degree sequence.

\paragraph{Community structure using the agents' labels.} In order to estimate the overlap between the communities detected using the \texttt{Infomap} algorithm and the group of influence defined by our agents' labels, we use the normalized mutual information coefficient $I_{\mathrm{norm}}$ \citep{danon2005comparing}. 
We find that labels are actually able to reproduce the community structures of collaboration networks coming from both the economic and the scientific domains. 
$I_{\mathrm{norm}}(\mathrm{Labels,~Infomap~clusters}) = 0.887 \pm 0.003$ for the ``Pharmaceuticals'' R\&D network, and $I_{\mathrm{norm}}(\mathrm{Labels,~Infomap~clusters}) = 0.952 \pm 0.002$ for the co-authorship network in interdisciplinary physics.
This result is even more remarkable if we consider that the \texttt{Infomap} algorithm detects structural clusters based on the probability flow of random walks in the network, while our label propagation mechanism consists of an assignment of a fixed membership attribute -- which is not only closer to a real phenomenon, but also computationally easier.

\subsection{Distribution of path lengths at link formation}
\label{sec:repr-distr-path}

\begin{figure}[htbp]
\begin{center}
\footnotesize{
(a) \includegraphics[width=0.38\textwidth,angle=0]{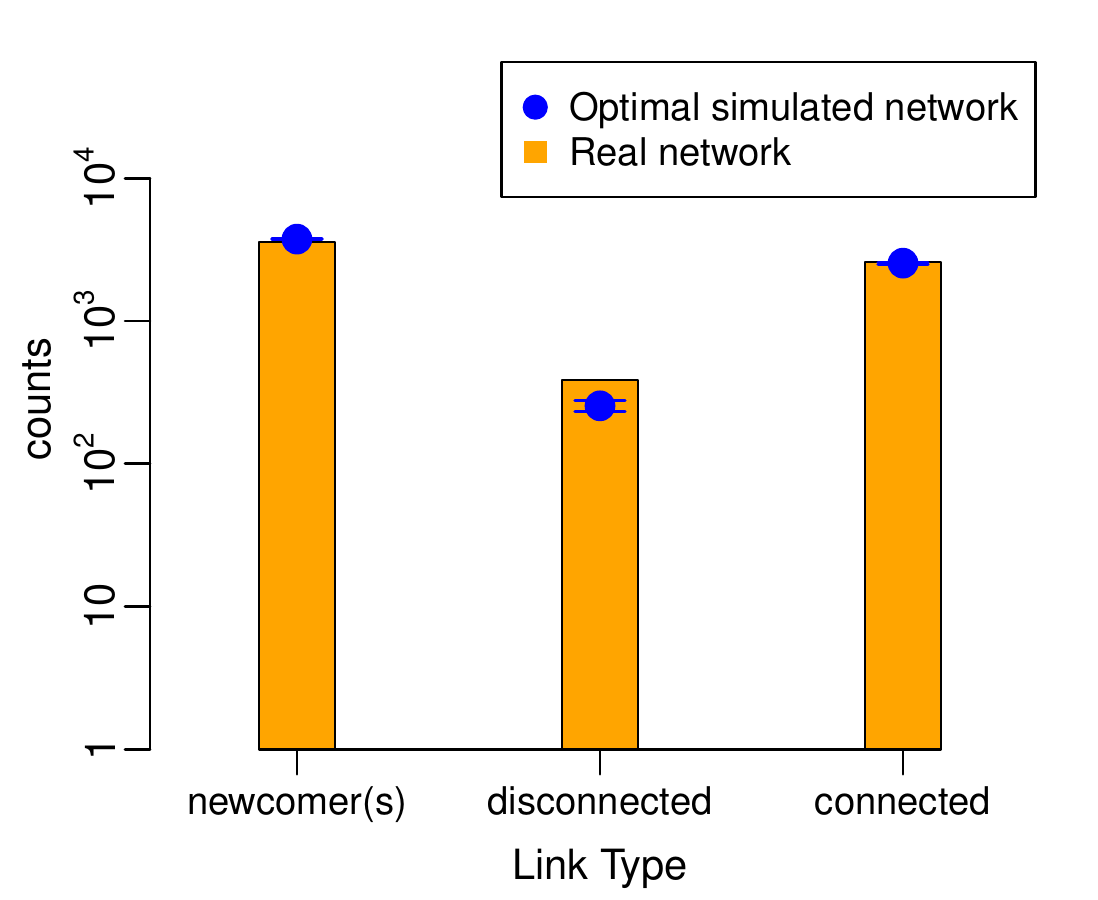}
(b) \includegraphics[width=0.38\textwidth,angle=0]{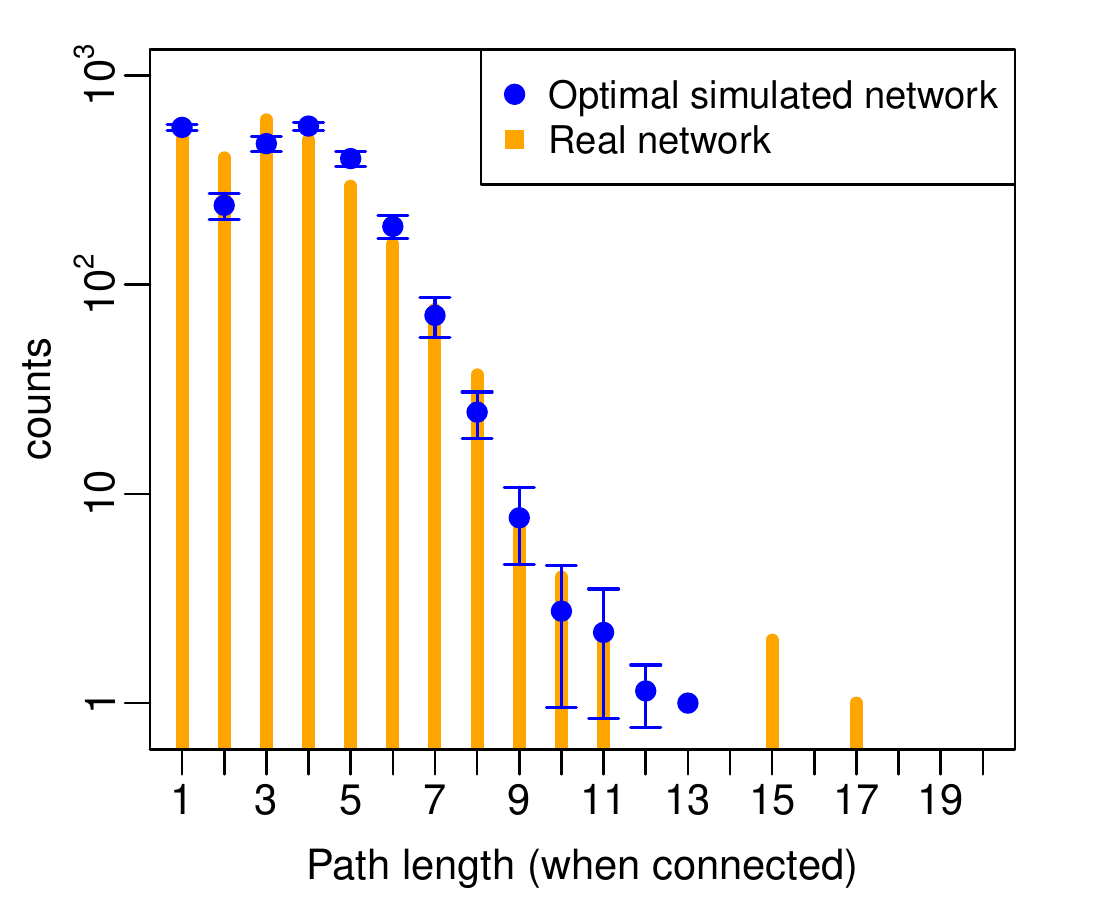}
}
\end{center}
\vspace{-8pt}
\caption[Temporal path length analysis in the Pharmaceuticals sectoral R\&D network.]{Temporal path length analysis for ``Pharmaceuticals'' R\&D network (SIC code 283) . (a) Distribution of link types for empirical and simulated networks: ``newcomer(s)'' means that at least one of the agents was isolated (i.e. not yet part of the network) before the link formation; ``disconnected'' refers to agents already belonging to the network, but placed in two disconnected components; ``connected'' refers to agents already belonging to the same network component prior to the link formation. (b) Distribution of path lengths at the moment of link formation (only for agents belonging to the same connected component).}
\label{fig:dynamic_path_lengths_pharma}
\end{figure}

\begin{figure}[htbp]
\begin{center}
\footnotesize{
(a) \includegraphics[width=0.38\textwidth,angle=0]{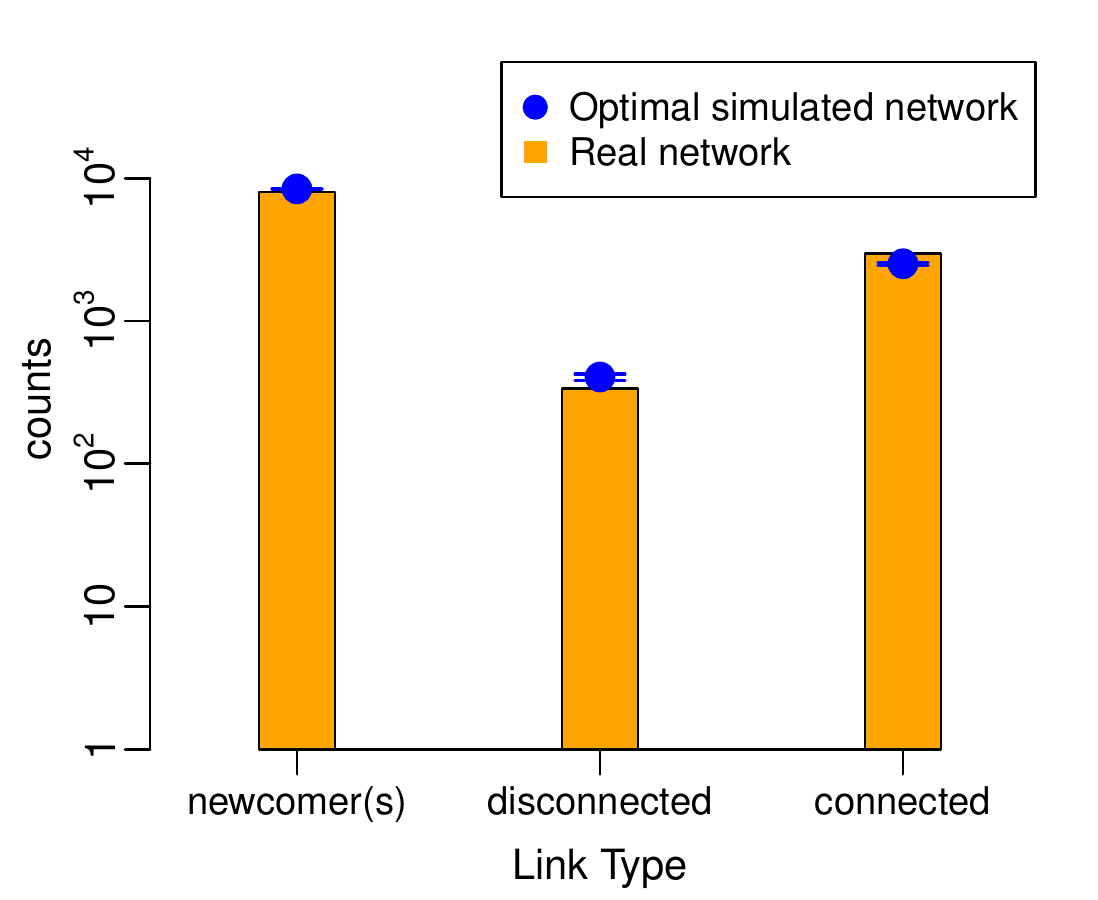}
(b) \includegraphics[width=0.38\textwidth,angle=0]{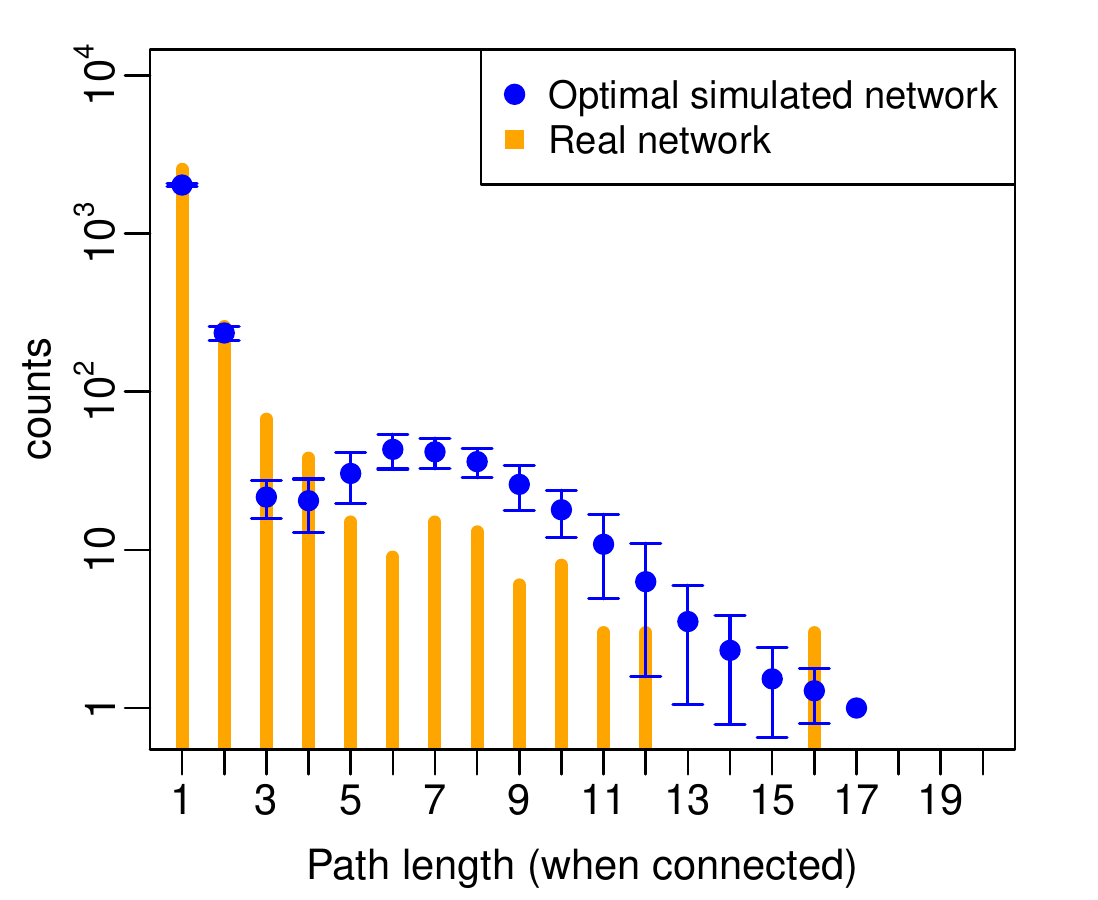}
}
\end{center}
\vspace{-8pt}
\caption[Temporal path length analysis for the co-authorship network in applied and interdisciplinary physics (PACS number 89).]{Temporal path length analysis for the co-authorship network in applied and interdisciplinary physics (PACS number 89) . (a) Distribution of link types for empirical and simulated networks: ``newcomer(s)'' means that at least one of the agents was isolated (i.e. not yet part of the network) before the link formation; ``disconnected'' refers to agents already belonging to the network, but placed in two disconnected components; ``connected'' refers to agents already belonging to the same network component prior to the link formation. (b) Distribution of path lengths at the moment of link formation (only for agents belonging to the same connected component).}
\label{fig:dynamic_path_lengths_PACS89}
\end{figure}

Finally, we compare the empirical and the simulated networks with respect to the distribution of path lengths between every pair of agents \emph{at the moment preceding the link formation}. 
This is different from the distribution of path lengths analyzed before, which was computed on the \emph{time-aggregated} networks. 
Now we are interested to know whether agents preferably form links with agents already part of the same connected component or with agents from another component or with newcomers. 
The respective distribution of link types is shown in Figure  \ref{fig:dynamic_path_lengths_pharma} for the ``Pharmaceuticals'' R\&D network, and in Figure  \ref{fig:dynamic_path_lengths_PACS89} for the co-authorship network in interdisciplinary physics.
In all cases, there is a higher number of links with agents inside the same connected component or with newcomers. 
We emphasize the very good match between the empirical and the simulated frequencies of link types. 

For links connecting agents which are already in the same connected component we can further discuss the network distance, or path length between two agents. 
It is interesting whether agents at larger network distances are still able to know each other and to form a link. 
Trivially, agents at distance 1 have already a collaboration (and can start a new one), whereas agents at distance 2 have one collaborator in common. 
We report our findings about the path length between agents before they engage in a collaboration 
in Figure  \ref{fig:dynamic_path_lengths_pharma} for the ``Pharmaceuticals'' R\&D network, and in Figure  \ref{fig:dynamic_path_lengths_PACS89} for the co-authorship network in interdisciplinary physics.
We see that in the case of R\&D networks agents preferably choose close collaborators for a new collaboration (path length up to 5), whereas for co-authorship networks agents prefer previous collaborators or collaborators at distance 2. 
In conclusion, the model correctly predicts the formation of links between agents no matter whether they are already in the same network component or not and gives an exact calculation of the shortest path length at the moment of link formation.

\section{Discussion and Conclusion}

\paragraph{Commonalities in collaboration networks.}

In the present paper, we have explored the structure and dynamics of collaboration networks in two different domains, 
{R\&D alliances between firms and co-authorship relations between scientists.}
Despite their different origin, these collaboration networks share a number of common features that can be even found on the {sub-domain} level (SIC and PACS numbers). 
These empirical features include the right-skewed distribution of collaboration sizes (Figure  \ref{fig:size_collab_events}), the distribution of activities to engage in a collaboration (Figure  \ref{fig:activities_rep})  which are very stable across domains and over time, the pronounced community structure of the networks and the existence of a giant-connected component (Figure  \ref{fig:COAUTnet89_empirical_simulated_communities}).

These commonalities motivated us to use the same agent-based model to explain the structure and dynamics of these collaboration networks. 
Precisely, we {have compared} the outcome on the \emph{systemic level}, i.e. the networks simulated by the agent-based model and the observed networks, to conclude whether our assumptions for the interactions on the \emph{agent level} are justified. 
We remark that reproducing systemic features along very different dimensions indeed lends evidence {to the validity of our agent-based model,} because it cannot simply be obtained by a fitting procedure. 
Specifically, our model is able to reproduce the distributions of degree, of path length, of local clustering coefficients, of component sizes and of path lengths between every pair of agents at the moment of link formation, without imposing any constraints on these features during the calibration procedure.

\paragraph{Strategies of agents choosing collaboration partners.}

The agent-based model builds on five probabilities to form a link with another agent, which depend on the label of the initiator (newcomer vs.  established agent) and on the counterparty (newcomer vs. established agent with the same or a different label). 
These agent-centric probabilities are calibrated using only three macroscopic features of the empirical networks (mean values of degree, path length and clustering coefficient).
Remarkably, we find that these probabilities {have very similar values}, regardless of the domains (R\&D networks vs co-authorship networks) and the {sub-domains}  (SIC and PACS numbers). 

Interpreting these probabilities as \emph{strategies} of an agent to choose a collaboration partner, we can {obtain} the following insights: 

(i)  For all R\&D and co-authorship networks, established agents prefer to form links with other established  agents ($p^{*L}_{s}+p^{*L}_{d} > 55\%$).

(ii)  When forming a link with an established agent, the initiator tends to select a counterparty with the same label, i.e. belonging to the same community ($p^{*L}_{s} \ge p^{*L}_{d}$). 
Comparing the two domains, we find that this general tendency is 10 times larger in co-authorship networks. 
The probability to select a co-author from  a \emph{different} community $p^{*L}_{d}$ equals the lowest possible value, 5\%, in all cases. 
 
(iii) A difference between domains is observed in the strategy of the newcomers. 
 For R\&D networks, newcomers tend to enter the network by forming links with \emph{established agents} ($p^{*N\!L}_{l} > p^{*N\!L}_{nl}$).
This finding is consistent with empirical evidence \citep{powell1996interorganizational, rosenkopf2008investigating}.
However, for  all co-authorship networks  newcomers tend to enter the network by forming links with other \emph{newcomers}
($p^{*N\!L}_{nl} > p^{*N\!L}_{l}$).
So, the fact that $p^{*N\!L}_{nl} \ge 0.55$ in co-authorship networks {clearly supports this hypothesis.}

The difference in the strategies of newcomers in R\&D and co-authorship networks can be attributed to the higher entry barriers in economic systems compared to academic environments. 
An exception from these general observations can be only found for one sectoral network ``R\&D, laboratory and testing'', where the strategies of newcomers  are more like in co-authorship networks. 
We attribute this deviation to the high technological dynamism in this sector.

\paragraph{Network-endogenous and -exogenous factors.}

Following the distinction in the literature \citep{rosenkopf2008investigating} we argue that the strategies of agents in choosing their collaboration partners are determined by {both} endogenous and exogenous factors. 
These are known to be crucial in the formation and evolution of the R\&D alliances \citep{rosenkopf2008investigating}. 
However, they have been usually considered separately by empirical and theoretical works \citep{garas2017selection, KogutWalkerShan1997, powell1996interorganizational, cowan2004network, BurtStrucHoles92}, and to our knowledge no study has analyzed their importance in co-authorship networks.

\emph{Network-endogenous} factors {cover the information that the initiator has} about the network, for instance information about the network position (i.e. social capital) of its potential partners.
Thus, these factors take {into} account collaboration patterns already present  in the networks.
{These factors are captured} by the probabilities to link to  a \emph{labeled agent}, $p^{L}_{s}$,  $p^{L}_{d}$ and $p^{N\!L}_{l}$.
\emph{Network-exogenous} factors do not consider such information, but instead use 
external information such as the technological, scientific or geographical proximity of the agents.
These factors are {captured} by the probabilities to link to a \emph{newcomer}, $p^{L}_{n}$ and $p^{N\!L}_{nl}$.

Comparing the two {types} of factors, we find that \emph{network-endogenous} factors are predominant in the formation of new collaborations {in each of the collaboration networks analyzed in this study. 
In other words, the existing network structures explain most of the newly formed links. 
In terms of linking probabilities, this means that} $p^{*L}_{s}$+$p^{*L}_{d}$+$p^{*N\!L}_{l}$ is always bigger than $p^{*L}_{nl}$+$p^{*N\!L}_{nl}$ (where $^{*}$ refers to the optimal probability) for  all sectoral R\&D networks and co-authorship networks. 
This result is also in line with the empirical finding \citep{gulati1995social, podolny1993status} that firms in R\&D networks prefer to establish alliances with other firms which have an history of previous alliances.

\paragraph{Reconstruction of communities by means of labels.}
In our model, labels represent the fact that agents belong to certain communities. 
This way, newcomers and established agents can be distinguished. 
Moreover, different labels allow to further differentiate between groups of agents with a certain interest. 
The label dynamics explained in Section \ref{sec:agent-based-model} provides a mechanism of label propagation. 

We point out that our assumption about the label attribute is in agreement with the results reported by \citet{yang2012ground_truth_communities}, that have identified the presence of communities based on ground truth in real networks. 
Such communities include nodes that do not necessarily share features such as the same geographical provenience, or the belonging to the same institution. 
They are rather defined dynamically, through consecutive interactions and link formation. 
The same reasoning holds for both R\&D and co-authorship networks, where communities of collaborating agents {do not depend} on their geographical or knowledge distance, but {are defined by} the subsequent propagation of a (virtual) membership attribute, which is the ``label''. 

It is remarkable that this rather abstract setup for labels is indeed able to reproduce  the distributions of communities present in the collaboration networks from both domains (see Figure  \ref{fig:distributions_of_communities}). 
The overlap in communities, measured through a normalized mutual information criterion, is around 90\% for all collaboration networks.
In Table \ref{table:collaboration_networks_clusters} in Appendix C, we have shown that such community structure cannot be expected at random from the degree sequence.  
Thus, we can conclude that labels represent a simple and elegant way to capture various network-endogenous factors which drive agents in both domains, R\&D collaborations and co-authorship networks, to form communities.
While the existence of communities is an empirical fact, the \emph{rules for their formation} are not fully understood. 
With this work, we provide evidence that such rules can be {inferred} from the empirical networks and are not only able to reproduce the community structure, but also other{, more sophisticated} features of the networks.

\bibliographystyle{sg-bibstyle} \bibliography{all_references_MT.bib}

\section*{Appendix A}

\subsection*{Empirical distribution of event sizes}
We report the distributions of partners per collaboration event for the two analyzed data sets. In the R\&D network, this quantity represents the number of firms per R\&D alliance, and in the co-authorship network the number of authors per paper.
Four representative distributions (two from each domain) were shown in Figure  \ref{fig:size_collab_events} in Section \ref{sec:data_and_methodology}.

\begin{figure}[htbp]
\begin{center}
\includegraphics[width=0.8\textwidth]{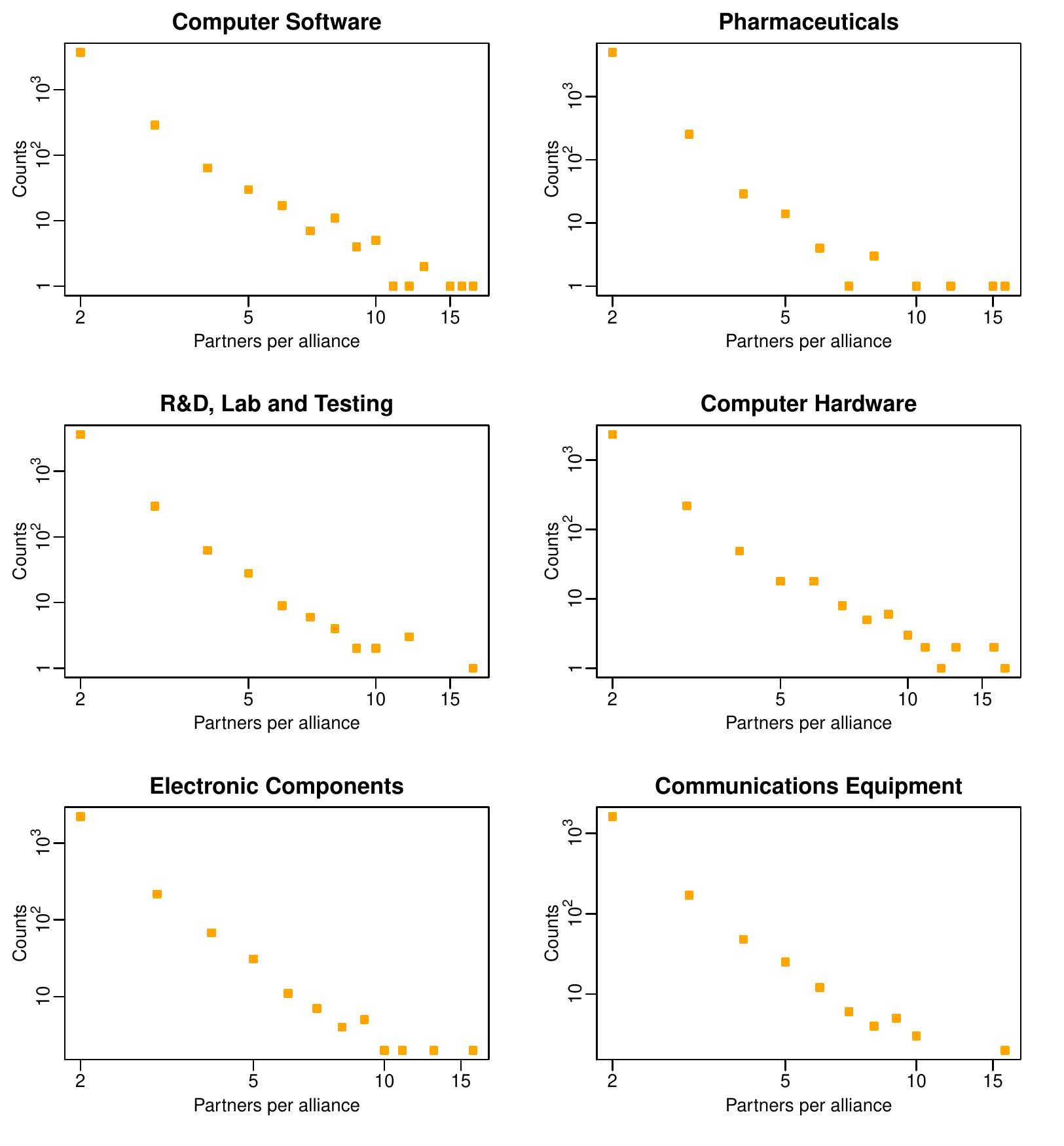} 
\end{center}
\vspace{-16pt}
\caption[Distribution of the number of partners per alliance for the six largest industrial sectors.]{Distribution of the number of partners per alliance for the six largest industrial sectors, as measured from the SDC data set.}
\label{fig:sectoral_partners}
\end{figure}

Most of the collaborations (93\%) are stipulated between two partners, but some alliances -- the so-called \textit{consortia} -- involve three or more partners.
These features are found also when considering separately the six largest industrial sectors with only small differences in the tails of the respective distributions.  
The plots of the distributions for the six largest industrial sectors are in Figure  \ref{fig:sectoral_partners}.

\begin{figure}[htbp]
\begin{center}
\includegraphics[width=0.8\textwidth]{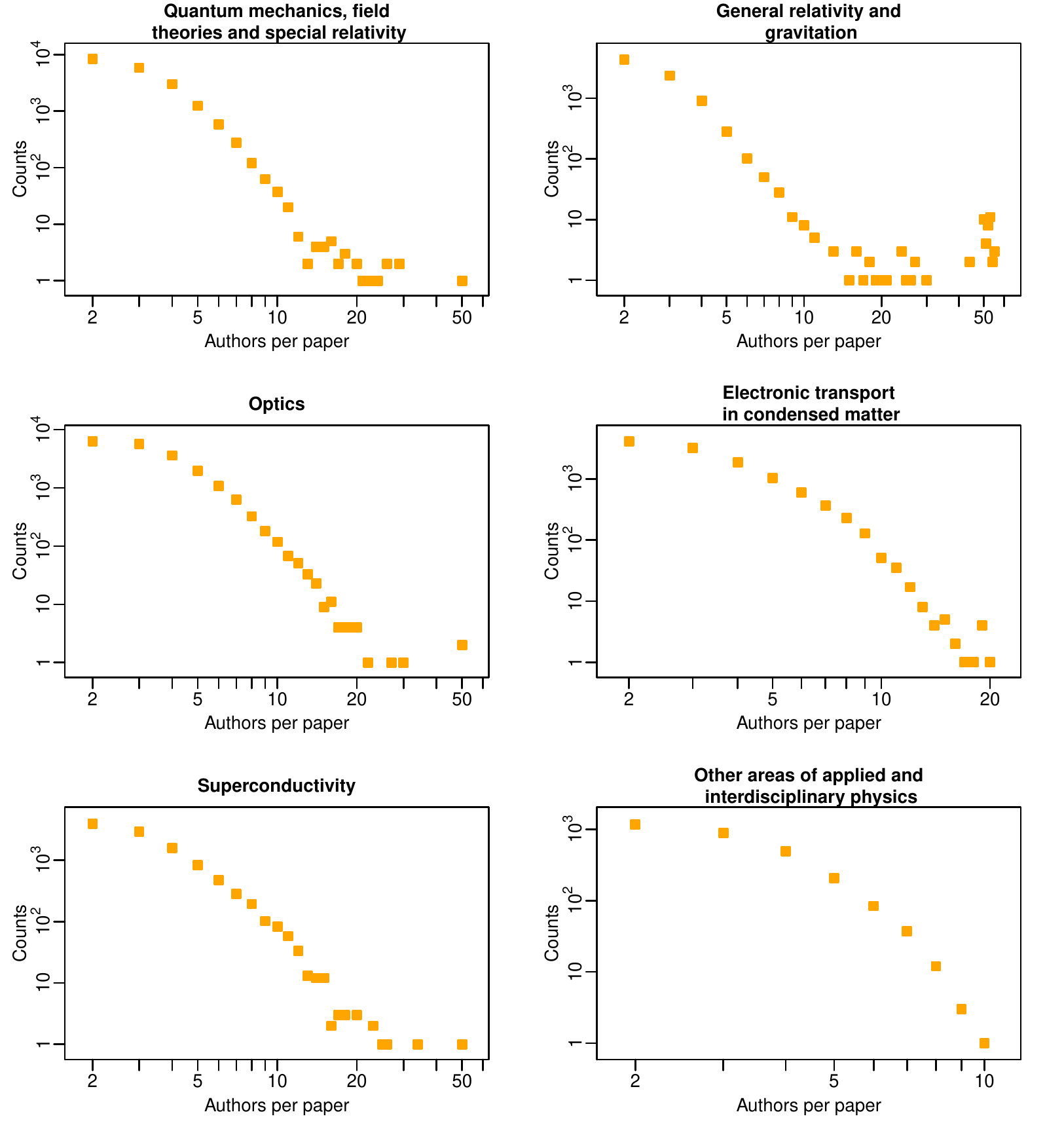} 
\end{center}
\vspace{-16pt}
\caption[Distribution of the number of authors per paper for our six representative co-authorship networks.]{Distribution of the number of authors per paper for our six representative co-authorship networks, as measured from the APS-MSAS data set.}
\label{fig:authors_per_paper_distr}
\end{figure}

In Figure  \ref{fig:authors_per_paper_distr} we report the size distribution of collaboration events for the different PACS number. 
Let us point out that in the General relativity and gravitation the observed strong increase of number of papers co-authored by about 50 people is an artifact of the data set.
As a matter of fact, we recognize that papers produced by large international collaborations, such as LIGO, may have many more than 50 co-authors, but their author lists have been cut to a maximum of 55 co-authors.
For most fields, this does not play any role since few papers are produced by such large collaborations.
PACS 04 (General relativity and gravitation) is an exception and we argue that this missing information makes the ABM unable to reproduce with good precision the network structure (see Section \ref{app:stats-obser-simul-nets} in Appendix C).

\subsection*{Empirical distribution of activities}
\begin{figure}[htbp]
\begin{center}
\includegraphics[width=0.95\textwidth]{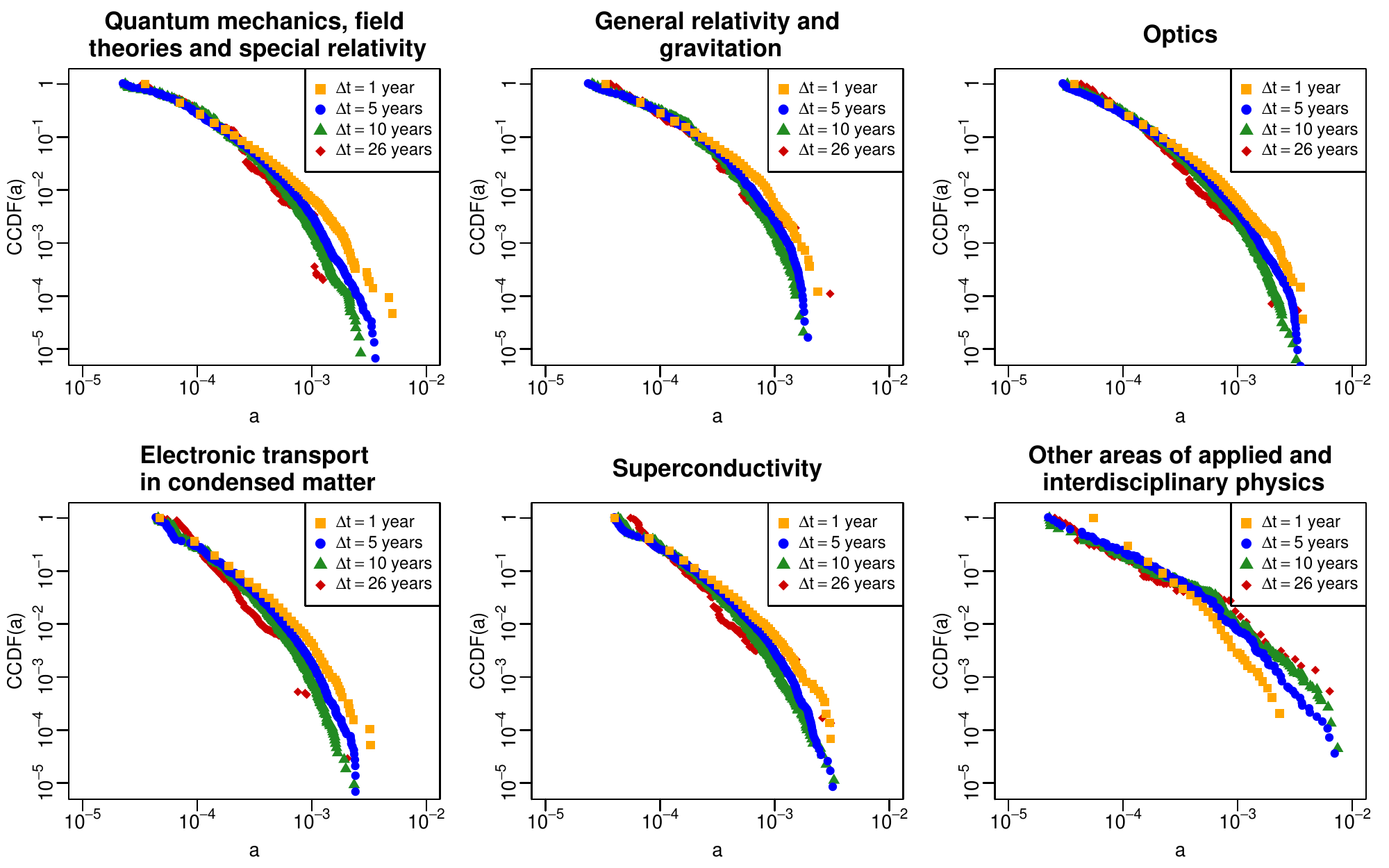}
\end{center}
\vspace{-16pt}
\caption[Agent activity distribution in the six representative co-authorship networks.]{Complementary cumulative distribution function (CCDF) of the empirical , measured for the six selected co-authorship networks in the APS-MSAS data set with 4 different time windows $\Delta t$ of 1, 5, 10 and 26 years. 
}
\label{fig:coauthorship_activities}
\end{figure}

We report the distribution of activities for all our representative co-authorship networks in  Figure  \ref{fig:coauthorship_activities}. 
As discussed in Section \ref{sec:data_and_methodology}, this distribution are not dependent of the chosen time window and always show a right-skewed distribution.
Note that the distribution of activities for the six sectoral R\&D networks are already reported in \citet{tomasello2014therole} (Supplementary information).

\section*{Appendix B}
\subsection*{Exploration of the parameter space}
In Section \ref{sec:impl-choos-optim}, we have discussed how we simulate the collaboration networks and how select the optimal set of probabilities from the simulations.
Here we would like to give some details about the simulations.
For each of the examined collaboration network, we explore the parameter space by varying the values of $p_s^{L}$,$p_d^{L}$ and $p_{nl}^{NL}$ between (0,1) by steps of 0.05. Since $p_s^{L}$ and $p_d^{L}$ are the probabilities of two mutually exclusive events, we also have to consider the condition $p_n^L=1-p_s^{L}-p_d^{L}>0$. 
This procedure gives $1/0.05-1=19$ values for $p_{nl}^{NL}$ and $(1/0.05-1)(1/0.05-2)/2=19*18/2$ combinations of values for $(p_s^{L},p_d^{L})$ creating a parameter space made of 3,249 points.
Thus, to explore the parameter space requires a remarkable computational effort because each of the 12 collaboration networks originates a parameter space composed of 3,249 points, for each of which we run 25 computer simulations -- for a total of around 1 million simulations. 

\subsection*{Average degree, path length and clustering coefficient for observed and optimal simulated networks}
\label{app:stats-obser-simul-nets}
In Table \ref{table:collaboration_networks_empirical}, we report the average degree $\mean{k}$, average path length $\mean{l}$ and global clustering coefficient $C$ for the empirical networks and for the simulated ones using the optimal set of probabilities. 
We also report the considered threshold.
It should be noted that -- given the extreme variability of the networks we test, in terms of size, density and modularity -- we are forced to adjust the error threshold value $\epsilon^0$ \citep[see][]{tomasello2014therole}, in order to find a meaningful number of parameter configurations that are able to reproduce the empirical network with a precision $\epsilon^0$. 
In particular for some co-authorship networks, we are not able to retrieve $\mean{k}$, $\mean{l}$ and $C$ with an accuracy as low as 2\% (which we could achieve for the time-aggregated R\&D network, \citep[see][]{tomasello2014therole}). 
However, all the values we obtain for our simulated networks are fairly accurate and deviate from the empirical values by less than 12\%.
 The only exception is represented by the co-authorship network in the field of general relativity and gravitation (PACS number 04), for which the model fails to generate a network matching all the three measures $\mean{k}$, $\mean{l}$ and $C$ at the same time.
 We argue that this is due to incomplete information in our data set and the consequent arising of a bimodal distribution of the number of partners per collaboration -- or, precisely, authors per paper -- in this scientific field. 
 Thus the linking probabilities and all the other results associated to this co-authorship network cannot be considered significant.
 
\begin{table}[htbp]
\scriptsize
\centering
\tabcolsep=0.14cm
\begin{tabular}{| l | r|r|r ||c|| r|r|r |}
\hline
\textbf{} & \boldmath$\mean{k}^{\mathrm{OBS}}$ & \boldmath$\mean{l}^{\mathrm{OBS}}$ & \boldmath$C^{\mathrm{OBS}}$ & $\epsilon^0$ & \boldmath$\mean{k}^*$ & \boldmath$\mean{l}^*$ & \boldmath$C^{*}$ \\
\hline
Aggregated R\&D network & 2.74 & 5.41 & 0.101 & 2\% & 2.76 & 5.33 & 0.098 \\
\hline
\textbf{Sectoral R\&D networks}  &  &  &  &  &  &  & \\ 
Pharmaceuticals (SIC 283) & 3.14 & 4.94 & 0.097 & 2\% & 3.13 & 4.95 & 0.097 \\ 
Computer hardware (SIC 357) & 5.12 & 3.70 & 0.161 & 6\% & 5.37 & 3.59 & 0.175  \\ 
Communications equipment (SIC 366) & 4.81 & 3.75 & 0.203 & 2\% & 4.83 & 3.76 & 0.210 \\ 
Electronic components (SIC 367) & 4.65 & 3.80 & 0.168 & 2\% & 4.76 & 3.83 & 0.174\\ 
Computer software (SIC 737) & 3.47 & 4.33 & 0.138 & 3\% & 3.56 & 4.27 & 0.141 \\ 
R\&D, laboratory and testing (SIC 873) & 3.37 & 5.15 & 0.205 & 3\% & 3.30 & 5.22 & 0.200 \\ 
\hline
\textbf{Co-authorship networks}  &  &  &  &  &  &  & \\ 
Quantum mechanics, field theories, special relativity (PACS 03)  & 5.22 & 6.43 & 0.379 & 12\% & 5.83 & 5.58 & 0.392  \\ 
General relativity and gravitation (PACS 04) & 7.84 & 6.27 & 0.666 & $\mathit{>30\%}$ & \textit{16.64} & \textit{4.39} & \textit{0.535} \\ 
Optics (PACS 42) & 6.92 & 6.40 & 0.425 & 10\% & 7.60 & 5.79 & 0.451 \\ 
Electronic transport in condensed matter (PACS 72) & 5.73 & 7.06 & 0.448 & 8\% & 6.15 & 6.58 & 0.471 \\ 
Superconductivity (PACS 74) & 7.05 & 5.87 & 0.443 & 7\% & 7.51 & 5.51 & 0.465 \\ 
Other areas of applied and interdisciplinary physics (PACS 89) & 3.60 & 8.28 & 0.462 & 8\% & 3.82 & 7.82 & 0.501\\ 
\hline
\end{tabular}
\caption[Summary of average statistics for the empirical and simulated collaboration networks.]{Summary of average statistics for the empirical and optimal simulated networks. For the empirical collaboration networks, we report the average degree, $\mean{k}^{\mathrm{OBS}}$, average path length, $\mean{l}^{\mathrm{OBS}}$, and global clustering coefficient, $C^{\mathrm{OBS}}$. For optimal simulated network network, we report the mean values over the 25 network realizations of average degree, $\mean{k}^*$, average path length, $\mean{l}^*$ and global clustering coefficient, $C^{*}$. We also report the error threshold or accuracy $\epsilon^0$. }
\label{table:collaboration_networks_empirical}
\end{table}

\section*{Appendix C}
\subsection*{Modularity for the empirical collaboration networks}
In Table \ref{table:collaboration_networks_clusters}, we report the number of communities detected by \texttt{Infomap} on the empirical networks and the normalized modularity score $Q$ for the empirical networks given the \texttt{Infomap} partitions. 
These values should be compared to the normalized modularity score $Q^{\mathrm{rand}}$ obtained from a set of 100 randomly generated networks using the degree sequence from the empirical networks. On each of the random network we have detected cluster of nodes using \texttt{Infomap} and computed the normalized modularity. 
Thus, $Q^{\mathrm{rand}}$s reported in \ref{table:collaboration_networks_clusters} are the mean normalized modularity scores from the 100 randomly generated networks for each sub-domain with their respective variance.
As discussed in Section \ref{sec:comm-str-and-groups-of-infl}, the modularity scores of the empirical networks are always higher than the ones coming from the randomly generated networks indicating that the detected modular structure is not an artifact of the degree sequence.
\begin{table}[htbp]
\centering
\tabcolsep=0.14cm
\begin{tabular}{| l || r|r|r|}
\hline

\textbf{} &  \textit{Clusters} & $Q$ & $Q^{\mathrm{rand}}$ \\
\hline
Aggregated R\&D network & 3,561 & 0.679 & 0.570 $\pm$ 0.001 \\
\hline
\textbf{Sectoral R\&D networks} &  &  & \\ 
Pharmaceuticals (SIC 283) & 860 & 0.607 & 0.438 $\pm$ 0.002 \\ 
Computer hardware (SIC 357) & 783 & 0.623 & 0.502 $\pm$ 0.002 \\ 
Communications equipment (SIC 366) & 749 & 0.653 & 0.461 $\pm$ 0.002 \\ 
Electronic components (SIC 367) & 302 & 0.502 & 0.311 $\pm$ 0.002 \\ 
Computer software (SIC 737) & 354 & 0.531 & 0.333 $\pm$ 0.002 \\ 
R\&D, laboratory and testing (SIC 873) & 256 & 0.527 & 0.317 $\pm$ 0.003 \\ 
\hline
\textbf{Co-authorship networks} &  &  & \\ 
Quant. mech., field theories, spec. relativity (PACS 03) & 3,029 & 0.779 & 0.2344 $\pm$ 0.0004 \\ 
General relativity and gravitation (PACS 04) & 1,207 & 0.795 & 0.128 $\pm$ 0.016 \\ 
Optics (PACS 42) & 2,853 & 0.794 & 0.195 $\pm$ 0.002 \\ 
Electronic transport in condensed matter (PACS 72) & 2,411 & 0.832 & 0.2609 $\pm$ 0.0004 \\ 
Superconductivity (PACS 74) & 1,663 & 0.769 & 0.208 $\pm$ 0.003 \\ 
Other applied and interdisciplin. physics (PACS 89) & 966 & 0.920 & 0.395 $\pm$ 0.001 \\ 
\hline
\end{tabular}
\caption[Modular properties of all examined collaboration networks.]{Modular properties for the aggregated R\&D network, the six largest sectoral R\&D networks, and the six representative co-authorship networks. For all domains, we consider the respective cumulative networks, i.e. the networks obtained by keeping all the links at any time. For each network, we report the number of clusters detected by the \texttt{Infomap} algorithm, the modularity score $Q$ of the network, and (as robustness check) the modularity score $Q^{\mathrm{rand}}$ obtained in a set of 100 randomly generated networks with the same size and degree sequence as the network under examination.}
\label{table:collaboration_networks_clusters}
\end{table}

\end{document}